\documentclass[%
final,         
a4paper,       
UKenglish,     
headsepline,   
abstracton,    
DIV10          
]{scrartcl}
\usepackage[utf8]{inputenc}
\usepackage[UKenglish]{babel}
\usepackage[T1]{fontenc}
\usepackage{textcomp}

\usepackage[verbose]{hyperref}  
\usepackage{url}                
\usepackage[sumlimits, intlimits, namelimits]{amsmath} 
\usepackage{amssymb}            
\usepackage[caption = true]{subfig} 
\usepackage{array}              
\usepackage{booktabs}           
\usepackage[longnamesfirst,round]{natbib} 
\usepackage{color}              
\usepackage{tikz}               
\usepackage[nogin]{Sweave}      

\hypersetup{
  bookmarksopen=true, 
  breaklinks=true,
  pdftitle={Hyper-g Priors for GLMs},
  pdfauthor={Daniel Saban\'es Bov\'e, Leonhard Held},
  pdfsubject={Statistics},
  pdfkeywords={Statistics, Modelling, Generalized Linear Models, %
    Bayesian Inference, Statistical Modelling, Model Choice, g-prior,
    hyper-g} 
}

\usepackage{xifthen}            
\usepackage{array}
\usepackage{amssymb}

\newcommand{\blanco}[1]{  } 

\newcommand{\latin}[1]{\textit{#1}}

\newcommand{\abk}[1]{\mbox{#1}\xdot}
\DeclareRobustCommand\xdot{\futurelet\token\Xdot}
\def\Xdot{\ifx\token\bgroup.\else\ifx\token\egroup.\else
  \ifx\token\/.\else\ifx\token\ .\else\ifx\token!.\else
  \ifx\token,.\else\ifx\token:.\else\ifx\token;.\else
  \ifx\token?.\else\ifx\token/.\else\ifx\token'.\else
  \ifx\token).\else\ifx\token-.\else\ifx\token+.\else
  \ifx\token~.\else
  \ifx\token.\else.\ \fi\fi\fi\fi\fi\fi\fi\fi\fi\fi\fi\fi\fi\fi\fi\fi}

\newcommand{\eg}{\abk{e.\,g}}   
\newcommand{\ie}{\abk{i.\,e}}
\newcommand{\cf}{\abk{cf}}

\newcolumntype{L}[1]{>{$}p{#1}<{$}} 
\newcolumntype{M}{>{$}l<{$}}    
\newcolumntype{N}{>{$}c<{$}}    

{\section*{Acknowledgments}}%
{}

\DeclareMathOperator{\Bin}{Bin} 
\DeclareMathOperator{\Nor}{N} 
  
\DeclareMathOperator{\IG}{IG} 
\DeclareMathOperator{\Var}{Var} 
\DeclareMathOperator{\E}{\mathbb{E}} 
\DeclareMathOperator{\diag}{diag} 
\renewcommand{\P}{\operatorname{\mathbb{P}}} 

\newcommand{\R}{\mathbb{R}}

\newcommand{\ml}[2][]{
  \ifthenelse{\isempty{#1}}%
  {\widehat{#2}_{\scriptscriptstyle{ML}}}%
  {\widehat{#2}^{#1}_{\scriptscriptstyle{ML}}}
}

\newcommand{\given}{\,\vert\,} 

\newcommand{\abs}[1]{\left\lvert#1\right\rvert} 
\newcommand{\norm}[1]{\left\lVert#1\right\rVert} 

\newcommand{\partialv}[3][]{%
  \ifthenelse{\isempty{#1}}
  {\frac{\partial\,#2}{\partial\,#3}}
  {\frac{\partial^{#1} #2}{\partial\,#3^{#1}}} 
} 

\newcommand{\partials}[3][]{%
  \ifthenelse{\isempty{#1}}
  {\frac{d\,#2}{d\,#3}}
  {\frac{d^{#1} #2}{d\,#3^{#1}}}
} 

\newcommand{\dseps}[2][]{%
  \ifthenelse{\isempty{#1}}
  {\frac{d}{d\,#2}}
  {\frac{d^{#1}}{d\,#2^{#1}}}
}

\newcommand{\dsepv}[2][]{%
  \ifthenelse{\isempty{#1}}
  {\frac{\partial\,}{\partial\,#2}}
  {\frac{\partial^{#1}}{\partial\,#2^{#1}}}
}

\def\frutiger{\usefont{T1}{pfr}{l}{sc}\selectfont}

\newcommand{\uzhlogo}{
  \begin{figure}[!h]
    \begin{center}
      \centerline{\includegraphics[width=5cm]{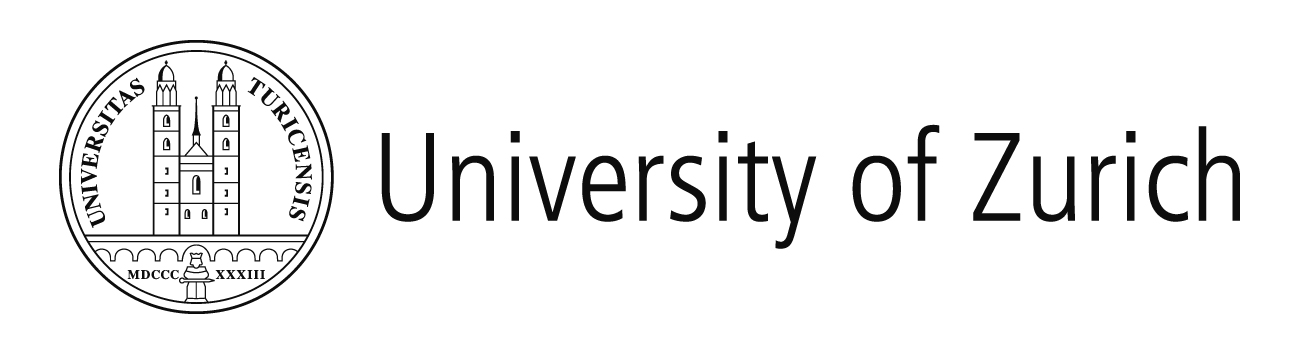}}
    \end{center}
  \end{figure}
}

\blanco{
\makeatletter
\@removefromreset{equation}{chapter}
\renewcommand\theequation{\@arabic\c@equation}
\makeatother
}

\begin{document}

\title{{\frutiger Hyper-$g$ Priors for Generalized Linear Models}}
\author{%
  {\frutiger Daniel Saban\'es Bov\'e}%
  \footnote{E-mail: \href{mailto:daniel.sabanesbove@ifspm.uzh.ch}{\url{daniel.sabanesbove@ifspm.uzh.ch}}}
  \and
  {\frutiger Leonhard Held}%
  \footnote{E-mail: \href{mailto:leonhard.held@ifspm.uzh.ch}{\url{leonhard.held@ifspm.uzh.ch}}}  
}
\publishers{\uzhlogo \vspace*{-1cm} {\frutiger Institute of Social and Preventive Medicine, Biostatistics Unit}} 
\date{{\frutiger Version: \today}}
\maketitle

\begin{abstract}
  \noindent We develop an extension of the classical Zellner's $g$-prior to
  generalized linear models. The prior on the hyperparameter~$g$ is handled in a
  flexible way, so that any continuous proper hyperprior $f(g)$ can be used,
  giving rise to a large class of hyper-$g$ priors. Connections with the
  literature are described in detail. A fast and accurate integrated Laplace
  approximation of the marginal likelihood makes inference in large model spaces
  feasible. For posterior parameter estimation we propose an efficient and
  tuning-free Metropolis-Hastings sampler. The methodology is illustrated with
  variable selection and automatic covariate transformation in the Pima Indians
  diabetes data set.
\end{abstract}

\noindent \emph{Keywords}: 
$g$-prior, 
generalized linear model, 
integrated Laplace approximation,
variable selection, 
fractional polynomials.

\section{Introduction}
\label{sec:introduction}


Assume that we have observed responses $y_{i}$ coming from a generalized linear
model (GLM, see \citealp{McCullaghNelder1989}) incorporating the covariate
vectors $\boldsymbol{x}_{i} \in \R^{p}$ via the linear predictors $\eta_{i} =
\beta_{0} + \boldsymbol{x}_{i}^{T}\boldsymbol{\beta}$, $i=1, \dotsc, n$. The
response function (inverse link function) $h$ transforms $\eta_{i}$ to the mean
$\E(y_{i}) = \mu_{i} = h(\eta_{i})$, which in turn is mapped to the canonical
parameter
$\theta_{i} = (db/d\theta)^{-1}(\mu_{i})$ 
of the exponential family.
Here 
$db/d\theta$
is the first derivative of the function $b$ that defines the
form of the likelihood for $\boldsymbol{y} = (y_{1}, \dotsc, y_{n})^{T}$ via
\begin{equation}
  \label{eq:glm-likelihood}
  f(\boldsymbol{y} \given \beta_{0}, \boldsymbol{\beta})
  \propto
  \exp
  \left\{
    \sum_{i=1}^{n}
    \frac{y_{i} \theta_{i} - b(\theta_{i})}{\phi_{i}}
  \right\},
\end{equation}
where each $\theta_{i}$ depends on the intercept $\beta_{0}$ and the vector
$\boldsymbol{\beta}$ of regression coefficients as described above. Often the
canonical response function 
$h = db/d\theta$
is used which leads to the identity
$\theta_{i} = \eta_{i} = \beta_{0} + \boldsymbol{x}_{i}^{T}\boldsymbol{\beta}$.
The dispersions $\phi_{i} = \phi / w_{i}$ are assumed known and can incorporate
weights $w_{i}$. The variance 
$\Var(y_{i}) = \phi_{i} d^{2}b/d\theta^{2}(\theta_{i})$ 
is expressed through the variance function 
$v(\mu_{i}) = d^{2}b/d\theta^{2}((db/d\theta)^{-1}(\mu_{i}))$ 
as $\Var(y_{i}) = \phi_{i} v(\mu_{i})$.


A Bayesian analysis starts by assigning prior distributions to the unknown model
parameters $\beta_{0}$ and $\boldsymbol{\beta}$. However, usually there is not
only uncertainty with respect to the model parameters, but also to the model
itself, see \eg \citet{ClydeGeorge2004}. Let $\gamma$ be the model index
contained in some model space $\Gamma$. Typically, the variable selection
problem is considered, where $\gamma \in \{0, 1\}^{m}$ contains binary inclusion
indicators for all $m$ available covariates. Here we think more generally of
uncertainty about the form (including the dimension $p_{\gamma}$) of the
covariate vectors $\boldsymbol{x}_{\gamma i}$, which may also comprise different
transformations of the original variables. For example, when $\gamma$ indicates
a quadratic transformation of $x_{i}$, then $\boldsymbol{x}_{\gamma i} = (x_{i},
x_{i}^{2})^{T}$. Thus, priors $f(\beta_{0}, \boldsymbol{\beta}_{\gamma} \given
\gamma)$ need to be assigned, for all models $\gamma \in \Gamma$. Manual
elicitation of all these priors is clearly infeasible when $\Gamma$ is large. In
this situation priors which automatically derive from $\gamma$ are attractive,
and we will propose such a prior in this paper. Model inference then uses the
posterior model probabilities
\begin{equation}
  \label{eq:post-model-probs}
  f(\gamma \given \boldsymbol{y}) 
  \propto
  f(\boldsymbol{y} \given \gamma) f(\gamma),
  \quad
  \gamma \in \Gamma,
\end{equation}
which combine the marginal likelihood
\begin{equation}
  \label{eq:marginal-likelihood}
  f(\boldsymbol{y} \given \gamma) = 
  \int_{\R^{p_{\gamma} + 1}} 
  f(\boldsymbol{y} \given \beta_{0}, \boldsymbol{\beta}_{\gamma}, \gamma) 
  f(\beta_{0}, \boldsymbol{\beta}_{\gamma} \given \gamma)
  \,d\beta_{0} d\boldsymbol{\beta}_{\gamma}
\end{equation}
with the prior model probabilities $f(\gamma)$. 


In the special case of the classical normal linear model with known error
variance $\phi$ and $w_{i} \equiv 1$, the $g$-prior for the regression
coefficients was proposed by \citet{Zellner1986} as a ``reference informative
prior''. It is the normal distribution with zero mean vector and covariance
matrix $g \phi (\boldsymbol{X}_{\gamma}^{T}\boldsymbol{X}_{\gamma})^{-1}$,
\begin{equation}
  \label{eq:standard-g-prior}
  \boldsymbol{\beta}_{\gamma} \given g, \phi \sim 
  \Nor_{p_{\gamma}}\bigl(
  \boldsymbol{0}_{p_{\gamma}}, 
  g \phi (\boldsymbol{X}_{\gamma}^{T}\boldsymbol{X}_{\gamma})^{-1}
  \bigr),
\end{equation}
and is usually combined with a locally uniform (Jeffreys) prior on $\beta_{0}$,
assuming that the design matrix $\boldsymbol{X}_{\gamma} =
(\boldsymbol{x}_{\gamma 1}, \dotsc, \boldsymbol{x}_{\gamma n})^{T}$ has been
centered to ensure $\boldsymbol{X}_{\gamma}^{T}\boldsymbol{1}_{n} =
\boldsymbol{0}_{p_{\gamma}}$. Often also the error variance $\phi$ is assumed
unknown and assigned a Jeffreys prior.
The $g$-prior can be interpreted as the conditional posterior of
$\boldsymbol{\beta}_{\gamma}$ given a locally uniform prior and an imaginary
sample $\boldsymbol{y}_{0} = \boldsymbol{0}_{n}$ from the normal linear model
with original design matrix $\boldsymbol{X}_{\gamma}$ and scaled error variance
$g\phi$. This implements the idea that after accounting for the mean level
$\beta_{0}$ not included in the $g$-prior, there is no difference between
observations due to the covariates in $\boldsymbol{X}_{\gamma}$ modelled through
$\boldsymbol{\beta}_{\gamma}$.
In addition to this nice interpretation, the $g$-prior has other advantages,
such as invariance of the implied prior for the linear predictor under rescaling
and translation of the covariates \citep[p.~71]{RobertSaleh1991}, and automatic
adaption to situations with near-collinearity between different covariates
\citep[p.~193]{Robert2001}.


The hyperparameter $g > 0$ in~\eqref{eq:standard-g-prior} acts as an inverse
prior sample size, and is thus very sensitive to prior elicitation. Larger
values of $g$ lead to preference of less complex models, a phenomenon known as
the Lindley-Jeffreys paradox (\citealp{Lindley1957}; see also
\citealp[p.~161]{RobertChopinRousseau2009}). Moreover, a fixed $g$ does not
allow the Bayes factor of a perfectly fitting model versus the null model go to
infinity \citep{BergerPericchi2001}. Therefore, much research has been done in
developing automatic specifications of $g$ \citep{GeorgeFoster2000,
  HansenYu2001, FernandezLeySteel2001, CuiGeorge2008}. The multivariate Cauchy
priors of \citet{ZellnerSiow1980} correspond to fully Bayesian inference with an
inverse-gamma prior for $g$. Unfortunately, the corresponding marginal
likelihood $f(\boldsymbol{y} \given \gamma)$ has no closed form. Therefore
\citet{LiangPauloMolinaClydeBerger2008} proposed the hyper-$g$ prior, which is a
special case of the incomplete inverse-gamma prior by \citet{CuiGeorge2008}.
These hyperpriors retain a closed form expression for $f(\boldsymbol{y} \given
\gamma)$ which is vital for efficient model inference.


In this article we develop an extension of the classical
$g$-prior~\eqref{eq:standard-g-prior} to GLMs. The prior on the
hyperparameter~$g$ is handled in a flexible way, so that any continuous
proper hyperprior $f(g)$ can be used. In
Section~\ref{sec:generalized-hyper-g-prior}, this generalized hyper-$g$ prior is
derived and connections with the literature are described. Because model
inference is the main practical use of this automatic prior formulation, we will
propose a fast and accurate numerical approximation of the marginal likelihood
in Section~\ref{sec:implementation}. Section~\ref{sec:implementation} also
covers posterior parameter estimation with a tuning-free Markov chain Monte
Carlo (MCMC) sampler. The methodology is applied to variable selection in
Section~\ref{sec:variable-selection} and to fractional polynomial modelling in
Section~\ref{sec:fractional-polynomials}. Section~\ref{sec:discussion} discusses
possibilities for future research.

\section{The generalized hyper-$g$ prior}
\label{sec:generalized-hyper-g-prior}

\subsection{Prior construction}
\label{sec:generalized-hyper-g-prior:construction}

Consider the imaginary sample $\boldsymbol{y}_{0} = h(0)\boldsymbol{1}_{n}$ from
the GLM with original design matrix $\boldsymbol{X}_{\gamma}$ and weights vector
$\boldsymbol{w} = (w_{1}, \dotsc, w_{n})^{T}$, but scaled dispersion $g\phi$.
Using an improper flat prior for the regression coefficients vector
$\boldsymbol{\beta}_{\gamma}$, its posterior given $\boldsymbol{y}_{0}$ is
proportional to the likelihood~\eqref{eq:glm-likelihood},
\begin{equation}
  \label{eq:imag-posterior}
  f(\boldsymbol{\beta}_{\gamma} \given \boldsymbol{y}_{0}, g, \gamma) 
  \propto
  \exp
  \left\{
    \frac{1}{g\phi}
    \sum_{i=1}^{n}
    h(0) w_{i}\theta_{i} - w_{i}b(\theta_{i})
  \right\}.
\end{equation}
This distribution can be recognized as the \citet[formula 2.6]{ChenIbrahim2003}
prior, although the authors have only considered the case with $w_{i} \equiv 1$.
Similar to their theorem~3.1, we can prove that the mode of this distribution is
at $\boldsymbol{\beta}_{\gamma} = \boldsymbol{0}_{p_{\gamma}}$ (see
Appendix~\ref{sec:prior-mode-zero-proof}). Moreover, it results from standard
Bayesian asymptotic theory (\eg \citealp{BernardoSmith2000}, p.~287)
that this distribution converges for $n\to \infty$ to the normal distribution
\begin{equation}
  \label{eq:generalized-g-prior}
  \boldsymbol{\beta}_{\gamma} \given g, \gamma \sim 
  \Nor_{p_{\gamma}}\bigl(
  \boldsymbol{0}_{p_{\gamma}}, 
  g\phi c
  (\boldsymbol{X}_{\gamma}^{T}\boldsymbol{W}\boldsymbol{X}_{\gamma})^{-1}
  \bigr)  
\end{equation}
where 
$c = v(h(0)) \cdot dh/d\eta(0)^{-2}$ and $\boldsymbol{W} =
\diag(\boldsymbol{w})$, because the inverse of the expected Fisher information
$I(\boldsymbol{\beta}_{\gamma})$ evaluated at the mode is
$I(\boldsymbol{0}_{p_{\gamma}})^{-1} = g\phi c
(\boldsymbol{X}_{\gamma}^{T}\boldsymbol{W}\boldsymbol{X}_{\gamma})^{-1}$
\citep[\cf][theorem~2.3]{ChenIbrahim2003}. The ``generalized
$g$-prior''~\eqref{eq:generalized-g-prior} differs from the standard
$g$-prior~\eqref{eq:standard-g-prior} only by the constant~$c$ and the weight
matrix~$\boldsymbol{W}$. Both are especially important in binomial regression
when $w_{i}$ is the sample size of the observed proportion, say $y_{i} = s_{i} /
w_{i}$ if $s_{i} \sim \Bin(w_{i}, \mu_{i})$ is the number of successes: In
Table~\ref{tab:distributions} it can be seen that only for the Bernoulli family
$c \neq 1$.
\begin{table}
  \centering
  \begin{tabular}{llc}
    \toprule
    Family      & Link                  & $c$ \\
    \midrule
    Gaussian    & Identity              & $1$ \\
                & (Log)                 & $1$ \\
    Poisson     & Log                   & $1$ \\
                & Identity              & ($0$) \\
    Bernoulli   & Logit                 & $4$ \\
                & Cauchit               & $\pi^{2}/4$ \\ 
                & Probit                & $\pi/2$ \\
                & Complementary log-log & $e - 1$ \\
    Gamma       & Log                   & $1$ \\
    Inverse Gaussian & (Log)            & $1$ \\
    \bottomrule
  \end{tabular}
  \caption{Exponential families, usual link functions and resulting factors $c$.
    Note that for the Gamma and the Inverse Gaussian family, the natural
    links $\mu^{-1}$ and $\mu^{-2}$, respectively, cannot be used because then
    $h(0) = \infty$. Parenthesized links should not be used because the
    uniqueness of the prior mode at $\boldsymbol{\beta}_{\gamma} = \boldsymbol{0}_{p_{\gamma}}$ is
    not sure \citep{Wedderburn1976}. Parenthesized $c$'s point out problems
    there.}  
  \label{tab:distributions}
\end{table}

Since the intercept $\beta_{0}$ parametrizes the average linear predictor in
each model, we can use the improper flat prior $f(\beta_{0}) \propto 1$. Thus,
our generalized $g$-prior does not shrink the intercept towards zero, while the
prior on the regression coefficients again implements the non-informative prior
idea that $\boldsymbol{X}_{\gamma}$ has \latin{a priori} no effect on the
centered observations. The factor $g$ is assigned a (continuous) hyperprior
$f(g)$, which we treat generally in the paper. The hyperprior $f(g)$ used in our
approach must be proper to ensure that Bayes factor comparisons with the null
model, which does not include the parameter $g$, are valid. As $g$ is assigned a
hyperprior, we call the resulting prior a ``generalized hyper-$g$ prior''.

\subsection{Comparison with the literature}
\label{sec:generalized-hyper-g-prior:comparison}

%
%
An immediate question is why we do not use the exact \citet{ChenIbrahim2003}
prior, which is also a generalization of the standard $g$-prior. The main
problem with this conjugate prior given in~\eqref{eq:imag-posterior} is that it
does not have a closed form for non-normal exponential families, \ie the
normalizing constant of~\eqref{eq:imag-posterior} is unknown. This complicates
the computation of the marginal likelihood and the MCMC sampling considerably.
\citet{ChenHuangIbrahimKim2008} propose a solution where they run an MCMC
sampler on the full model, and then derive estimates for submodels. However,
this approach is not applicable in problems with simultaneous variable selection
and transformation as that presented in Section~\ref{sec:fractional-polynomials}
because no full model exists in that case.
Regarding the hyperparameter~$g$, \citet{ChenIbrahim2003} propose to assign an
inverse-gamma prior to it.

%
Alternatively, \citet{GuptaIbrahim2009} proposed the information matrix prior,
which uses the expected Fisher information matrix
$I(\boldsymbol{\beta}_{\gamma})$ similarly to a precision matrix for a normal
distribution up to a scalar variance factor~$g$:
\begin{equation}
  \label{eq:gupta-ibrahim}
  f_{GI}(\boldsymbol{\beta}_{\gamma} \given g, \gamma) \propto
  \abs{I(\boldsymbol{\beta}_{\gamma})}^{1/2}
  \exp
  \left\{
    -\frac{1}{2g}\boldsymbol{\beta}_{\gamma}^{T} I(\boldsymbol{\beta}_{\gamma})
    \boldsymbol{\beta}_{\gamma} 
  \right\}.
\end{equation}
This will only be a Gaussian distribution if the matrix
$I(\boldsymbol{\beta}_{\gamma})$ actually does not depend on
$\boldsymbol{\beta}_{\gamma}$, \eg for the normal linear model where the
standard $g$-prior is reproduced by~\eqref{eq:gupta-ibrahim}. By contrast, the
precision of our generalized $g$-prior in~\eqref{eq:generalized-g-prior} results
from evaluating $I(\boldsymbol{\beta}_{\gamma})$ at the prior mode, producing a
matrix which does not depend on $\boldsymbol{\beta}_{\gamma}$.
\citet{GuptaIbrahim2009} fix the hyperparameter $g$ at a ``moderately large''
value ($g \geq 1$) and do not consider inference for it.

%
The information matrix prior is strongly linked with the unit information prior
approach of~\citet{KassWasserman1995}, who proposed the general idea that the
amount of information in the prior on $\boldsymbol{\beta}_{\gamma}$ should be
equal to the amount of information about it contained in one observational unit.
The amount of information is measured by the (expected) Fisher information, so
that the precision is chosen as $n^{-1} I(\boldsymbol{0}_{p_{\gamma}})$ in the
normal prior
\begin{equation}
  \label{eq:kass-wasserman}
  f_{KW}(\boldsymbol{\beta}_{\gamma} \given g, \gamma) 
  =
  \Nor_{p_{\gamma}}
  \left(
    \boldsymbol{\beta}_{\gamma} \given 
    \boldsymbol{0}_{p_{\gamma}},
    n I(\boldsymbol{0}_{p_{\gamma}})^{-1}
  \right).
\end{equation}
This proposal is close to ours in~\eqref{eq:generalized-g-prior}, except that
the hyperparameter is fixed at $g=n$. Note that \citet{KassWasserman1995} also
required the nuisance parameter $\beta_{0}$ to be null-orthogonal to the
parameter of interest $\boldsymbol{\beta}_{\gamma}$, which we ensure by
centering the covariates around zero. The unit information prior was used by
\citet{NtzoufrasDellaportasForster2003} and \citet{OverstallForster2010} in the
GLM context.

%
\citet[p.~156]{HansenYu2003} also use the expected Fisher information, but
evaluate it at the maximum likelihood (ML) estimate
$\hat{\boldsymbol{\beta}}_{\gamma}$ to obtain a prior precision matrix:
\begin{equation}
  \label{eq:hansen-yu}
  f_{HY}(\boldsymbol{\beta}_{\gamma} \given g, \gamma)
  =
  \Nor_{p_{\gamma}}
  \left(
    \boldsymbol{\beta}_{\gamma}
    \given
    \boldsymbol{0}_{p_{\gamma}},
    g I(\hat{\boldsymbol{\beta}}_{\gamma})^{-1}
  \right).
\end{equation}
\citeauthor{HansenYu2003} find the dependence of their prior on the data
$\boldsymbol{y}$ ``hard to accept'', although it can be interpreted as an
empirical-Bayes approach. Also in this flavour, the authors maximize the
(approximate) conditional marginal likelihood in order to eliminate $g$.
Subsequent model selection is then based on a modified conditional marginal
likelihood (``minimum description length'').

%
Instead of using the \emph{expected} Fisher information matrix
$I(\boldsymbol{\beta}_{\gamma})$, \citet{WangGeorge2007} use the \emph{observed}
Fisher information matrix $J(\boldsymbol{\beta}_{\gamma})$. While for canonical
response functions the equality $I(\boldsymbol{\beta}_{\gamma}) =
J(\boldsymbol{\beta}_{\gamma})$ holds, in general
$J(\boldsymbol{\beta}_{\gamma})$ is different and depends on the observed
response vector. \citet{WangGeorge2007} evaluate it at the original response
$\boldsymbol{y}$ and the ML estimate to obtain the correlation structure of the
normal distribution:
\begin{equation}
  \label{eq:wang-george}
  f_{WG}(\boldsymbol{\beta}_{\gamma} \given g, \gamma)
  =
  \Nor_{p_{\gamma}}
  \left(
    \boldsymbol{\beta}_{\gamma} 
    \given
    \boldsymbol{0}_{p_{\gamma}}, 
    g J(\hat{\boldsymbol{\beta}}_{\gamma})^{-1} 
  \right).
\end{equation}
By comparison, our generalized $g$-prior~\eqref{eq:generalized-g-prior} does
not use the original data $\boldsymbol{y}$, but only the design matrix
$\boldsymbol{X}_{\gamma}$. Analogously to \citet{HansenYu2003},
\citet{WangGeorge2007} select model-specific values for $g$ by maximizing the
conditional marginal likelihood $f(\boldsymbol{y} \given g, \gamma)$, but they
also consider fully Bayesian inference for $g$.
 
%
\citet[p.~101]{MarinRobert2007} avoid the use of a Fisher information matrix
altogether when they propose the ``non-informative $g$-prior''
\begin{equation}
  \label{eq:marin-robert}
  f_{MR}(\boldsymbol{\beta}_{\gamma} \given g, \gamma) 
  =
  \Nor_{p_{\gamma}}
  \left(
    \boldsymbol{\beta}_{\gamma}
    \given
    \boldsymbol{0}_{p_{\gamma}},
    g   (\boldsymbol{X}_{\gamma}^{T}\boldsymbol{X}_{\gamma})^{-1}
  \right)
\end{equation}
for binary regression with probit and logit link functions. The factor $g$ is
assigned the improper prior $f(g) \propto g^{-3/4}$, which can be regarded as a
degenerate inverse-gamma distribution with shape $-1/4$ and scale $0$. Note that
using this improper hyperprior prohibits Bayes factor comparisons with the null
model.

\section{Implementation}
\label{sec:implementation}

\subsection{Marginal likelihood computation}
\label{sec:implementation:marg-likel-comp}


Under the generalized hyper-$g$ prior, the marginal likelihood introduced
in~\eqref{eq:marginal-likelihood} is
\begin{align}
  f(\boldsymbol{y} \given \gamma)
  &=
  \int_{\R^{p_{\gamma} + 1}} 
  f(\boldsymbol{y} \given \beta_{0}, \boldsymbol{\beta}_{\gamma}, \gamma) 
  \int_{\R_{+}} 
  f(\boldsymbol{\beta}_{\gamma} \given g, \gamma) f(g)
  \, dg
  \, d\beta_{0} d\boldsymbol{\beta}_{\gamma}
  \notag
  \\
  &=
  \int_{\R_{+}} 
  f(\boldsymbol{y} \given g, \gamma) f(g)
  \, dg,
  \label{eq:glm-marginal-likelihood}
\end{align}
where the conditional marginal likelihood of the GLM $\gamma$ (given $g$) is
\begin{equation}
  \label{eq:conditional-marg-lik}
  f(\boldsymbol{y} \given g, \gamma) 
  =
  \int_{\R^{p_{\gamma} + 1}} 
  f(\boldsymbol{y} \given \beta_{0}, \boldsymbol{\beta}_{\gamma}, \gamma)
  f(\boldsymbol{\beta}_{\gamma} \given g, \gamma)
  \, d\beta_{0} d\boldsymbol{\beta}_{\gamma}.
\end{equation}
Note that both \eqref{eq:conditional-marg-lik} and
\eqref{eq:glm-marginal-likelihood} are only defined up to a constant, which we
have fixed at unity, as we use the improper prior $f(\beta_{0}) \propto 1$. In
general, no closed form expressions are available. The obvious exception is the
special case of normal likelihood, which was mentioned in
Section~\ref{sec:introduction} and will be referred to again later on in this
section. Therefore, in order to be able to efficiently explore a large model
space $\Gamma$, we need to develop a fast but accurate numerical approximation
to the marginal likelihood. This will be a two-step procedure: The conditional
marginal likelihood~\eqref{eq:conditional-marg-lik} is computed by a Laplace
approximation. Plugging this into~\eqref{eq:glm-marginal-likelihood}, the
hyperparameter~$g$ will be integrated out with respect to its prior by numerical
integration. Together, this is an integrated Laplace approximation (ILA), which
was proposed more generally by~\citet{RueMartinoChopin2009}.


For ease of notation, denote by $\boldsymbol{\beta}_{0\gamma} = (\beta_{0},
\boldsymbol{\beta}_{\gamma}^{T})^{T}$ the vector of all coefficients. Then the
Laplace approximation \citep{Lindley1980,TierneyKadane1986}
of~\eqref{eq:conditional-marg-lik} is 
\begin{align}
  f(\boldsymbol{y} \given g, \gamma) 
  &\approx
  \frac
  {f(\boldsymbol{y} \given \boldsymbol{\beta}_{0\gamma}^{*}, \gamma)
    f(\boldsymbol{\beta}_{0\gamma}^{*} \given g, \gamma)}
  {\tilde{f}(\boldsymbol{\beta}_{0\gamma}^{*} \given \boldsymbol{y}, g, \gamma)} 
  \notag\\
  \begin{split}
    &= f(\boldsymbol{y} \given \boldsymbol{\beta}_{0\gamma}^{*}, \gamma)
    (2\pi g\phi c)^{-p/2}
    \abs{\boldsymbol{X}_{\gamma}^{T}\boldsymbol{W}\boldsymbol{X}_{\gamma}}^{1/2}
    \exp
    \left\{
      -\frac{1}{2} (g\phi c)^{-1} \boldsymbol{\beta}_{0\gamma}^{*T}
      \boldsymbol{X}_{\gamma}^{T}\boldsymbol{W}\boldsymbol{X}_{\gamma} 
      \boldsymbol{\beta}_{0\gamma}^{*}
    \right\}
    \\
    &\quad\times
    (2\pi)^{(p+1)/2} \abs{\boldsymbol{R}_{0\gamma}^{*}}^{-1/2}  
  \end{split}
  \label{eq:cond-marg-lik-approximation}
\end{align}
when $\tilde{f}(\boldsymbol{\beta}_{0\gamma} \given \boldsymbol{y}, g, \gamma)$
is the Gaussian approximation of the conditional coefficients posterior with
mean vector $\boldsymbol{\beta}_{0\gamma}^{*}$ and precision matrix
$\boldsymbol{R}_{0\gamma}^{*}$. Since the conditional coefficients prior can be
seen to have a normal kernel 
$f(\boldsymbol{\beta}_{0\gamma} \given g, \gamma)
\propto
\exp
\left\{
  -\frac{1}{2} 
  \boldsymbol{\beta}_{0\gamma}^{T}
  \boldsymbol{R}_{0\gamma}
  \boldsymbol{\beta}_{0\gamma}
\right\}$
with (singular) precision
\begin{equation}
  \label{eq:coefficients-normal-precision}
  \boldsymbol{R}_{0\gamma} = 
  \diag
  \left\{
    0,
    (g\phi c)^{-1} 
    \boldsymbol{X}_{\gamma}^{T}\boldsymbol{W}\boldsymbol{X}_{\gamma}
  \right\},
\end{equation}
the Bayesian iterative weighted least squares (IWLS) algorithm \citep{West1985,
  Gamerman1997} can be used to compute the moments of the Gaussian
approximation. Note that this is different and potentially more accurate than
the approach by \citet[p.~327]{RueMartinoChopin2009} who preserve the sparsity
of the prior precision $\boldsymbol{R}_{0\gamma}$ in the resulting posterior
precision $\boldsymbol{R}_{0\gamma}^{*}$. The accuracy of the Laplace
approximation~\eqref{eq:cond-marg-lik-approximation} can be even further
improved by including higher-order terms of the underlying Taylor expansion. For
canonical response functions, \citet{RaudenbushYangYosef2000} derived a
convenient correction factor corresponding to a sixth-order Laplace
approximation. In the applications of Sections~\ref{sec:variable-selection} and
\ref{sec:fractional-polynomials}, we have used this correction (see
Appendix~\ref{sec:higher-order-laplace} for details), which clearly improved the
ILA while requiring only slightly more computation time.

The one-dimensional integration in~\eqref{eq:glm-marginal-likelihood} is
performed on the log-scale over $z= \log(g)$ using Gauss-Hermite
quadrature. 
First, we find the (approximate) posterior mode $z^{*}$ and variance
$\sigma^{*2}$ of $z$ using its unnormalized (approximate) posterior density
\begin{equation}
  \label{eq:unnormalized-z-posterior}
  f(z, \boldsymbol{y} \given \gamma)
  = 
  f(\boldsymbol{y} \given z, \gamma) f(z).
\end{equation}
The mode $z^{*}$ is numerically determined by the \texttt{optimize} routine in
\texttt{R} \citep{R07, Brent1973}. The variance $\sigma^{*2}$ can be computed as
the negative inverse second derivative of the log posterior at $z^{*}$ by
numerical differentiation \citep[routine \texttt{dfridr}
from][p.~231]{PressTeukolskyVetterlingFlannery2007}.
Second, we apply the Gauss-Hermite quadrature \citep{NaylorSmith1982}
\begin{equation}
  \label{eq:gauss-hermite-approx}
  f(\boldsymbol{y} \given \gamma)
  \approx 
  \sum_{j=1}^{N} m_{j} f(z_{j}, \boldsymbol{y} \given \gamma), 
\end{equation}
where the actual weights $m_{j} = \omega_{j} \exp(t_{j}^{2}) \sqrt{2}
\sigma^{*}$ and nodes $z_{j} = z^{*} + \sqrt{2} \sigma^{*} t_{j}$ depend on
$z^{*}$, $\sigma^{*}$ as well as original weights $\omega_{j}$ and nodes
$t_{j}$, $j=1, \dotsc, N$. These can be obtained from the
\citet{GolubWelsch1969} algorithm, which is implemented in the
\texttt{R}-function \texttt{gauss.quad} \citep{Smyth2009}. $N=20$ seems to be
sufficient, given that this includes nodes in a range of about seven standard
deviations around $z^{*}$ (as then $\sqrt{2}t_{20} \approx 7.6$). Note that the
Gauss-Hermite approximation in~\eqref{eq:gauss-hermite-approx} is exact if $f(z,
\boldsymbol{y} \given \gamma)$ is the product of $\Nor(z \given z^{*},
\sigma^{*2})$ and a polynomial of at most order $2N - 1$.

\subsection{Metropolis-Hastings sampler}
\label{sec:implementation:mh-sampler}

Given a model $\gamma\in\Gamma$ we would like to sample from the joint posterior
of the model-specific parameters $\boldsymbol{\theta}_{\gamma} =
(\boldsymbol{\beta}_{0\gamma}^{T}, z)^{T}$. To this end, we propose a
tuning-free Metropolis-Hastings (MH) sampling scheme with proposal kernel
\begin{equation}
  \label{eq:mh-proposal-kernel}
  q(\boldsymbol{\theta}_{\gamma}' \given \boldsymbol{\theta}_{\gamma})
  =
  q(\boldsymbol{\beta}_{0\gamma}' \given z', \boldsymbol{\beta}_{0\gamma})
  q(z')
\end{equation}
for the proposal $\boldsymbol{\theta}_{\gamma}'$ given the current sample
$\boldsymbol{\theta}_{\gamma}$. 
The independence proposal density $q(z')$ is constructed by linearly
interpolating pairs $\bigl(z_{j}, f(z_{j}, \boldsymbol{y} \given \gamma)\bigr)$
and then normalizing this function to unity integral. Note that many pairs are
already available from the optimization and integration
of~\eqref{eq:unnormalized-z-posterior} in the marginal likelihood computation.
Thus, $q(z)$ is close to the posterior density $f(z \given \boldsymbol{y},
\gamma)$, suggesting high acceptance rates of the sampler. Also, generating
random variates from $q(z)$ using inverse sampling is straightforward as the
corresponding cumulative distribution function is piecewise quadratic.

For the coefficients, $q(\boldsymbol{\beta}_{0\gamma}' \given z',
\boldsymbol{\beta}_{0\gamma})$ is a Gaussian proposal density: Starting from the
current vector $\boldsymbol{\beta}_{0\gamma}$ and the proposed prior covariance
factor $g' = \exp(z')$, a single step of the Bayesian IWLS is made, resulting in
the mean vector 
and the precision matrix
of the proposal \citep{Gamerman1997}.
In order to compute the acceptance probability of the move from
$\boldsymbol{\theta}_{\gamma}$ to $\boldsymbol{\theta}_{\gamma}'$,
\begin{equation}
  \label{eq:mh-acceptance-probability}
  \alpha(\boldsymbol{\theta}_{\gamma}' \given \boldsymbol{\theta}_{\gamma})
  = 
  1 \wedge
  \frac
  {f(\boldsymbol{y} \given \boldsymbol{\beta}_{0\gamma}', \gamma)
    f(\boldsymbol{\theta}_{\gamma}' \given \gamma)} 
  {f(\boldsymbol{y} \given \boldsymbol{\beta}_{0\gamma}, \gamma)
    f(\boldsymbol{\theta}_{\gamma} \given \gamma)} 
  \cdot
  \frac
  {q(\boldsymbol{\theta}_{\gamma} \given \boldsymbol{\theta}_{\gamma}')}
  {q(\boldsymbol{\theta}_{\gamma}' \given \boldsymbol{\theta}_{\gamma})},
\end{equation}
note that the prior contributions have the form $f(\boldsymbol{\theta}_{\gamma}
\given \gamma) = f(\boldsymbol{\beta}_{\gamma} \given g, \gamma) f(g) g$, the
last factor $g$ being due to the change of variable $z = \log(g)$ in the
proposal parametrization. For the reverse proposal kernel value
$q(\boldsymbol{\theta}_{\gamma} \given \boldsymbol{\theta}_{\gamma}')$, another
IWLS step starting from the proposed vector $\boldsymbol{\beta}_{0\gamma}'$ and
the current factor $g = \exp(z)$ is necessary.

Besides producing parameter samples from the posterior
$f(\boldsymbol{\theta}_{\gamma} \given \boldsymbol{y}, \gamma)$, the
MH sampler can also be used to compute an MCMC estimate of the
marginal likelihood $f(\boldsymbol{y} \given \gamma)$, thereby providing an
independent check of the numerical estimate presented in
Section~\ref{sec:implementation:marg-likel-comp}. We will use the method by
\citet[section~2.1]{ChibJeliazkov2001}, which was competitive in a review
by~\citet{HanCarlin2001} and is still a benchmark for new developments
\citep[see \eg][]{NottKohnFielding2008}. The estimate is based on the basic
identity
\begin{equation}
  \label{eq:marg-lik-basic-identity}
  f(\boldsymbol{y} \given \gamma)
  =
  \frac
  {f(\boldsymbol{y} \given \boldsymbol{\theta}_{\gamma}^{*}, \gamma)
    f(\boldsymbol{\theta}_{\gamma}^{*} \given \gamma)}
  {f(\boldsymbol{\theta}_{\gamma}^{*} \given \boldsymbol{y}, \gamma)},
\end{equation}
where $\boldsymbol{\theta}_{\gamma}^{*}$ is usually chosen as a configuration
with high posterior density which is fixed before the MCMC sampling. Then the
detailed balance of the Markov chain ensures that the unknown posterior ordinate
can be estimated by
\begin{equation}
  \label{eq:posterior-ordinate-estimate}
  f(\boldsymbol{\theta}_{\gamma}^{*} \given \boldsymbol{y}, \gamma) 
  \approx
  \frac
  {\sum_{j=1}^{B} 
    \alpha(\boldsymbol{\theta}_{\gamma}^{*} \given
    \boldsymbol{\theta}_{\gamma}^{(j)})  
    q(\boldsymbol{\theta}_{\gamma}^{*} \given
    \boldsymbol{\theta}_{\gamma}^{(j)})} 
  {\sum_{k=1}^{B}
    \alpha(\boldsymbol{\theta}_{\gamma}^{(k)} \given 
    \boldsymbol{\theta}_{\gamma}^{*})},
\end{equation}
where the $\boldsymbol{\theta}_{\gamma}^{(j)}$ are the posterior samples and the
$\boldsymbol{\theta}_{\gamma}^{(k)}$ are iid draws from the proposal
distribution $q(\boldsymbol{\theta}_{\gamma} \given
\boldsymbol{\theta}_{\gamma}^{*})$. Since each acceptance probability
in~\eqref{eq:posterior-ordinate-estimate} requires two additional IWLS steps,
$4B$~additional IWLS steps are required for the \citet{ChibJeliazkov2001}
estimate if $B$~posterior samples are used.

\subsection{Performance in the conjugate case}
\label{sec:implementation:perf-conj-case}

For illustration of the performance of the proposed implementation, we consider
the special case of normal linear regression with fixed error variance $\phi$.
Using the $g$-prior~\eqref{eq:standard-g-prior}, the conditional coefficients
posterior is Gaussian,
\begin{equation}
  \label{eq:conjugate-cond-coefs-posterior}
  f(\boldsymbol{\beta}_{0\gamma} \given \boldsymbol{y}, g, \gamma)
  = 
  \Nor\left(
    \beta_{0} 
    \given 
    \bar{y}, \phi/n
    \right)
  \Nor
  \left(
    \boldsymbol{\beta}_{\gamma} 
    \given
    g(g+1)^{-1} \hat{\boldsymbol{\beta}}_{\gamma}, 
    \,
    g(g+1)^{-1} \phi (\boldsymbol{X}_{\gamma}^{T}\boldsymbol{X}_{\gamma})^{-1}
  \right),
\end{equation}
where the ordinary least squares estimate $\hat{\boldsymbol{\beta}}_{\gamma} =
(\boldsymbol{X}_{\gamma}^{T}\boldsymbol{X}_{\gamma})^{-1}
\boldsymbol{X}_{\gamma}^{T}\boldsymbol{y}$ is shrunk by the factor
$g(g+1)^{-1}$. Thus, the Laplace
approximation~\eqref{eq:cond-marg-lik-approximation} of the conditional marginal
likelihood is exact. It is given by
\begin{equation}
  \label{eq:conjugate-cond-marg-lik}
  f(\boldsymbol{y} \given g, \gamma)
  =
  (g+1)^{-p_{\gamma}/2}
  \exp
  \left\{
    (g+1)^{-1}
    \left[
      - \frac{SSR_{\gamma}}{2\phi}
    \right]
  \right\}
  \cdot
  \exp
  \left\{
    - \frac{SSE_{\gamma}}{2\phi}
  \right\},
\end{equation}
where $SSE_{\gamma}$ and $SSR_{\gamma}$ are the error and regression sums of
squares, respectively. From the form of~\eqref{eq:conjugate-cond-marg-lik} we
see that an inverse-gamma prior $\IG(a, b)$ on $g+1$ will be conjugate to this
likelihood. Since $g > 0$ must be ensured, this distribution must be truncated
to $(1, \infty)$, yielding the incomplete inverse-gamma prior
\citep[p.~891]{CuiGeorge2008}
\begin{equation}
  \label{eq:incomplete-inverse-gamma}
  f(g) = 
  M(a, b)
  (g+1)^{-(a+1)} \exp\{- b / (g+1) \} 
\end{equation}
with the normalising constant 
\begin{equation}
  \label{eq:iig-norm-constant}
  M(a, b) = \frac
  {b^{a}}
  {\int_{0}^{b} t^{a-1} \exp(-t)\, dt}
\end{equation}
and the corresponding marginal likelihood
\begin{equation}
  \label{eq:conjugate-iig-marg-lik}
  f(\boldsymbol{y} \given \gamma) 
  =
  \frac{M(a, b)}{M(a_{\gamma}, b_{\gamma})}
   \exp
  \left\{
    - \frac{SSE_{\gamma}}{2\phi}
  \right\},
\end{equation}
where the updated parameters $a_{\gamma} = a + p_{\gamma}/2$ and $b_{\gamma} =
SSR_{\gamma} / (2\phi) + b$ determine the posterior of $g$ in model~$\gamma$.

\begin{figure}[htbp]
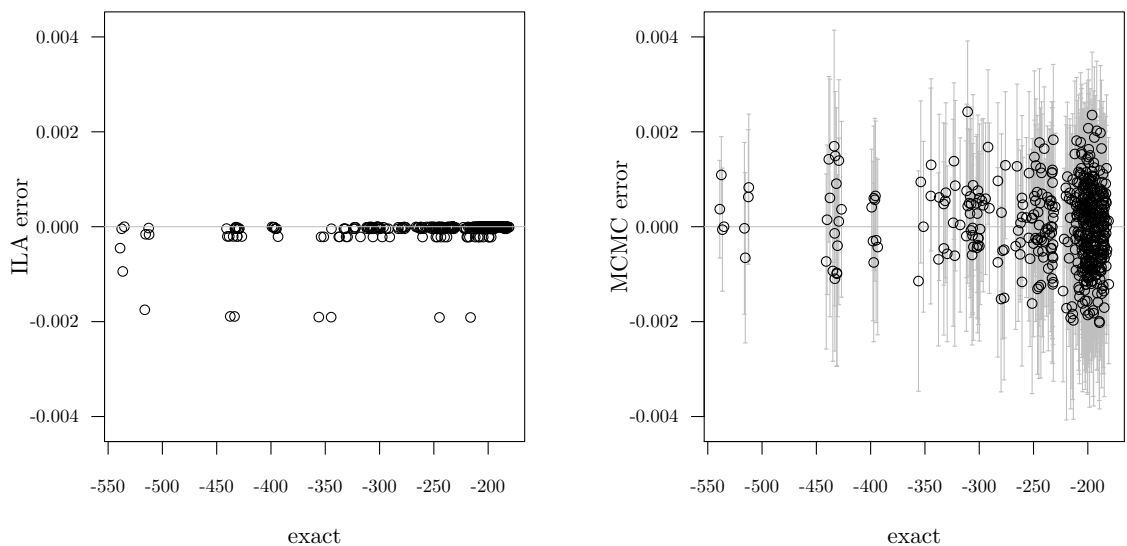
%
  \centering%
  \caption{Errors of ILA and MCMC estimates (y-axes) compared to the exact
    marginal likelihood values (x-axes) for all 512~models. MCMC estimates are
    based on $B=4500$ samples which were saved after
    burn-ins of length $1000$ (every
    2nd iteration). Note that the marginal likelihood
    values include the additional constant factor $\sqrt{2\pi\phi/n}$ compared
    to~\eqref{eq:conjugate-iig-marg-lik}.}%
  \label{fig:conjugate-marg-lik-comparison}%
\subfloat[Errors of the ILA estimates.%
\label{fig:conjugate-marg-lik-comparison:ila-error}]{%
\input{out-conj-ml-ila-error.tikz}
}%
\subfloat[Errors of the MCMC estimates. The vertical bars represent 95\%
MCMC confidence intervals (coverage is 95.1\% here).%
\label{fig:conjugate-marg-lik-comparison:mcmc-error}]{%
\input{out-conj-ml-mcmc-error.tikz}
}%
\end{figure}

In order to show results from a real data set, we consider the ozone data
introduced by \citet{BreimanFriedman1985} in the notation of
\citet{SabanesHeld2010}, where $n=330$. Deciding whether to
include each of the nine meteorological covariates $z_{0}$ and $z_{4}, \dotsc,
z_{11}$ in the linear regression of the daily maximum ozone concentration $y$
yields a model space $\Gamma$ of size $2^{9} = 512$. For all $\gamma \in
\Gamma$, the ILA~\eqref{eq:gauss-hermite-approx} and the MCMC
estimate~\eqref{eq:marg-lik-basic-identity} of the exact marginal likelihood
value~\eqref{eq:conjugate-iig-marg-lik} were computed using the fixed
(full-model variance estimate) $\phi = 19.75$ and the hyperprior
parameters $a=0.01, b=0.01$.
Figure~\ref{fig:conjugate-marg-lik-comparison} shows that the errors of ILA and
MCMC are very small here compared to the absolute true values.

\begin{figure}
  \centering
  \caption{True posterior density of $z$ (solid line) compared with the ILA
    (dashed line) and MCMC (histogram) estimates. Small ticks above the
    horizontal axis indicate where nodes $z_{j}$ for the construction of the ILA
    estimate $q(z)$ were located (\cf
    Section~\ref{sec:implementation:mh-sampler}).} 
  \label{fig:ozone-z-post-comparison}
\input{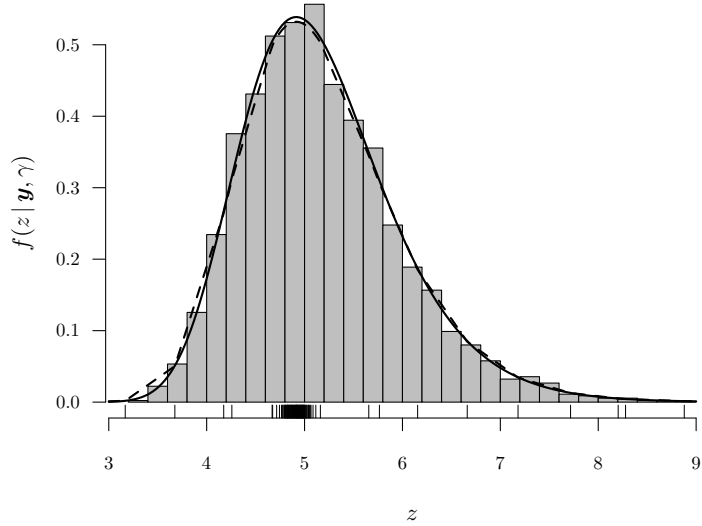}
\end{figure}

For all models, the acceptance rates of the MH algorithm were above
$97$\%.
Figure~\ref{fig:ozone-z-post-comparison} shows that even for the model with the
lowest acceptance rate, the true posterior density of $z=\log(g)$ is very close
to its ILA estimate $q(z)$. This explains the almost perfect acceptance rates of
the MH scheme.

\section{Variable selection}
\label{sec:variable-selection}

We illustrate the methodology for non-normal data with the Pima Indians diabetes
data set \citep{FrankAsuncion2010,Ripley1996}, which contains
$n=532$ complete records on diabetes presence and $m=7$ associated
covariates described in Table~\ref{tab:Pima-data-description}. First, we
restrict ourselves to variable selection in the logistic regression model,
yielding a model space $\Gamma$ of size~$2^{7} = 128$. In
Section~\ref{sec:fractional-polynomials}, we will also consider power
transformations of the covariates.

\begin{table}[htbp]
  \centering
  \begin{tabular}{ll}
    \toprule
    Variable & Description \\ 
    \midrule
    $y$ & Signs of diabetes according to WHO criteria (Yes = $1$, No = $0$) \\
    $x_{1}$ & Number of pregnancies \\
    $x_{2}$ & Plasma glucose concentration in an oral glucose tolerance
    test [mg/dl]\\
    $x_{3}$ & Diastolic blood pressure [mm Hg] \\
    $x_{4}$ & Triceps skin fold thickness [mm] \\
    $x_{5}$ & Body mass index (BMI) [kg/m\textsuperscript{2}]\\
    $x_{6}$ & Diabetes pedigree function \\
    $x_{7}$ & Age [years] \\    
    \bottomrule
  \end{tabular}
  \caption{Description of the variables in the Pima Indians diabetes data set.} 
  \label{tab:Pima-data-description}
\end{table}

Three different prior distributions for the covariance factor $g$ are compared
for a fully Bayesian analysis:
\begin{description}
\item[F1] $f(g) = \IG(g\given 1/2, n/2)$, corresponding to the
  \citet{ZellnerSiow1980} approach;
\item[F2] $f(g) = 1/n (1 + g/n)^{-2}$, corresponding to the hyper-$g/n$ prior
  \citep[p.~416]{LiangPauloMolinaClydeBerger2008}; 
\item[F3] $f(g) = \IG(g\given 0.001, 0.001)$, which is a standard choice for
  variance parameters.
\end{description}
We also consider model-specific empirical-Bayes estimation of $g$ using the
conditional marginal likelihood~\eqref{eq:conditional-marg-lik}, abbreviating
this approach as \textbf{\textsf{EB}}.
Moreover, the standard criteria \textbf{\textsf{AIC}} and \textbf{\textsf{BIC}}
are computed for each model.
We use the prior model probabilities
\begin{equation}
  \label{eq:mult-adjust-model-prior-probs}
  f(\gamma) = \frac{1}{m + 1} \binom{m}{p_{\gamma}}^{-1}
\end{equation}
for an appropriate multiplicity adjustment \citep{GeorgeMcCulloch1993,
  ScottBerger2010}. Posterior model probabilities then follow
from~\eqref{eq:post-model-probs}, where for EB the maximized conditional
marginal likelihood~\eqref{eq:conditional-marg-lik} and for BIC the
approximation $\exp(-1/2\, \mathrm{BIC})$ \citep[\eg][]{KassRaftery1995} is used
instead of $f(\boldsymbol{y} \given \gamma)$. Similar model weights proportional
to $\exp(-1/2\, \mathrm{AIC})$ can also be calculated for AIC as proposed by
\citet{BucklandBurnhamAugustin1997}.


In Table~\ref{tab:Pima-varselect-inclusion-probs}, the resulting posterior
probabilities and AIC weights for variable inclusion are shown. All methods
clearly select $x_{1}$, $x_{2}$, $x_{5}$ and $x_{6}$. The corresponding model is
the \latin{maximum a posteriori} (MAP) model in F1, F2, F3 and BIC, while for EB
and AIC also $x_{7}$ is included in the top model. This covariate would be
included as well in the median probability model \citep{BarbieriBerger2004} for
all methods except BIC. For $x_{3}$ and $x_{4}$, the evidence for inclusion is
consistently weak. For comparison, \citet{HolmesHeld2006} used vague iid normal
priors for all coefficients and a flat model prior $f(\gamma) = 2^{-7}$,
obtaining clear evidence for inclusion of the MAP covariates.

It is interesting that the inclusion probabilities under F1, F2 and F3 are
qualitatively similar. The reason could be that the sample size is relatively
large in this example, reducing the importance of the hyperprior specification
for $g$. For EB, most inclusion probabilities are even higher than for F3. The
AIC weights are more similar to F2 probabilities (except for $x_{7}$). The BIC
based probabilities are mostly lower, and close to the (not shown) probabilities
under F1 when a flat model prior is used.
 
\begin{table}
  \centering
%
\begin{tabular}{lrrrrrr}\toprule
\multicolumn{1}{l}{}&\multicolumn{1}{c}{F1}&\multicolumn{1}{c}{F2}&\multicolumn{1}{c}{F3}&\multicolumn{1}{c}{EB}&\multicolumn{1}{c}{AIC}&\multicolumn{1}{c}{BIC}\tabularnewline
\midrule
$x_{ 1 }$&$0.961$&$0.965$&$0.968$&$0.970$&$0.972$&$0.946$\tabularnewline
$x_{ 2 }$&$1.000$&$1.000$&$1.000$&$1.000$&$1.000$&$1.000$\tabularnewline
$x_{ 3 }$&$0.252$&$0.309$&$0.353$&$0.384$&$0.309$&$0.100$\tabularnewline
$x_{ 4 }$&$0.248$&$0.303$&$0.346$&$0.376$&$0.296$&$0.103$\tabularnewline
$x_{ 5 }$&$0.998$&$0.998$&$0.998$&$0.998$&$0.998$&$0.997$\tabularnewline
$x_{ 6 }$&$0.994$&$0.995$&$0.996$&$0.996$&$0.998$&$0.987$\tabularnewline
$x_{ 7 }$&$0.528$&$0.586$&$0.629$&$0.659$&$0.670$&$0.334$\tabularnewline
\bottomrule
\end{tabular}\caption{Posterior probabilities and AIC weights for variable inclusion in the
  Pima data.} 
\label{tab:Pima-varselect-inclusion-probs}
\end{table}

\begin{figure}
  \centering
  \caption{Model-averaged fitted probabilities in the Pima Indians variable
    selection example.}
  \label{fig:pima-varselect-bma-modelfits}
\input{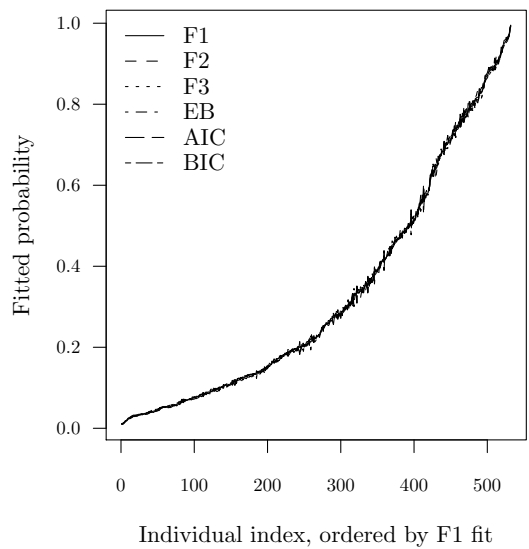}
\end{figure}

\begin{figure}
  \centering
  \caption{Errors of ILA with respect to MCMC estimates of the marginal likelihood
  under F3, for all 128 models in the Pima Indians variable selection example.
  MCMC estimates are based on (at least) $B=5000$ samples which were saved after
  burn-ins of length $1000$ (every
  2nd iteration). The vertical bars represent 95\% MCMC
  confidence intervals (coverage is 72.7\% here).}
  \label{fig:pima-varselect-f3-ila-errors}
\input{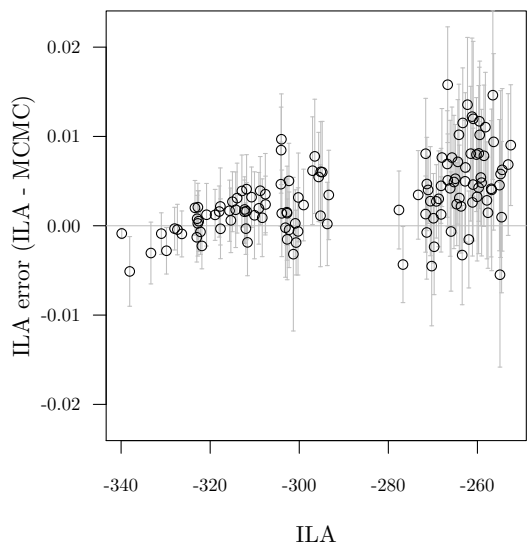}
\end{figure}

While the posterior inclusion probabilities are visibly different for the six
approaches, the model-averaged fits to the data are very close, as shown in
Figure~\ref{fig:pima-varselect-bma-modelfits}. In parallel to estimating the
posterior parameter distributions leading to these fitted probabilities for F1,
F2, F3 and EB, we also estimated the marginal likelihood by MCMC. The resulting
MCMC estimates were close to the ILA estimates, comparison plots looking like
Figure~\ref{fig:pima-varselect-f3-ila-errors} for F3. Note that the coverage of
the MCMC confidence intervals is lower than in
Figure~\ref{fig:conjugate-marg-lik-comparison:mcmc-error}, because the ILA
approximations are not exact.

\section{Fractional polynomials}
\label{sec:fractional-polynomials}

Fractional polynomials (FPs) are used for systematic power transformations of
the covariates $x_{1}, \dotsc, x_{m}$ \citep{RoystonAltman1994}. They widen the
class of ordinary polynomials insofar as the powers are taken from the fixed set
$\{-2, -1, -1/2, 0, 1/2, 1, 2, 3\}$, which also contains square roots,
reciprocals and the logarithm by the \citet{BoxTidwell1962} convention $x^{0}
\equiv \log(x)$. For each covariate $x_{k}$, at most two powers are chosen and
collected in the tuple $\boldsymbol{p}_{k}$, while the corresponding
coefficients are collected in the vector $\boldsymbol{\alpha}_{k}$, determining
the FP transform $x_{k}^{\boldsymbol{p}_{k}}\boldsymbol{\alpha}_{k}$. The
special case $p_{k1} = p_{k2}$ is handled by multiplication with the logarithm,
\eg $x_{k}^{(2, 2)} = \bigl(x_{k}^{2}, x_{k}^{2}\log(x_{k})\bigr)$. Variable
selection is embedded in this framework, because $x_{k}$ is not included in the
model if $\boldsymbol{p}_{k} = \emptyset$. Each model is thus uniquely
identified by $\gamma = (\boldsymbol{p}_{1}, \dotsc, \boldsymbol{p}_{m})$, the
covariate vectors are $\boldsymbol{x}_{\gamma i} = (x_{1i}^{\boldsymbol{p}_{1}},
\dotsc, x_{mi}^{\boldsymbol{p}_{m}})^{T}$ and the vector of regression
coefficients is $\boldsymbol{\beta}_{\gamma} = (\boldsymbol{\alpha}_{1}^{T},
\dotsc, \boldsymbol{\alpha}_{m}^{T})^{T}$. \citet{SabanesHeld2010} implemented
Bayesian inference in normal linear FP models, and more details on FPs can be
found in references therein.

The model space $\Gamma$ comprises $45^{m}$ models, and thus the use of an
automatic prior for the parameter $\boldsymbol{\beta}_{\gamma}$, conditional on
the model $\gamma$, is very attractive. The generalized
$g$-prior~\eqref{eq:generalized-g-prior} is automatic and only depends on the
global hyperparameter $g$. We will again compare the three fully Bayesian
approaches (F1, F2, F3) with the empirical-Bayes procedure (EB) which were
introduced in Section~\ref{sec:variable-selection} and avoid manual
specification of $g$.
The prior model probabilities $f(\gamma) = \prod_{k=1}^{m}f(\boldsymbol{p}_{k})$
depend on the prior FP transformation probabilities 
\begin{equation}
  f(\boldsymbol{p}_{k}) = 
  \frac{1}{3} 
  \binom{7 + \abs{\boldsymbol{p}_{k}}}{\abs{\boldsymbol{p}_{k}}}^{-1}
\end{equation}
which have the same form as~\eqref{eq:mult-adjust-model-prior-probs}: each
degree $\abs{\boldsymbol{p}_{k}} \in \{0, 1, 2\}$ is equally probable, and all
tuples $\boldsymbol{p}_{k}$ of the same degree are equally probable. This
implements Jeffreys's ``simplicity postulate'' that simpler models must have
greater prior probability than more complex models
\citep[section~1.6]{Jeffreys1961}, indeed the null model has the largest prior
probability $3^{-m}$.

For the Pima data the model space $\Gamma$ has size $45^{7} \approx
3.7\cdot 10^{11}$, rendering an
exhaustive evaluation of all $\gamma\in\Gamma$ infeasible. Therefore we use an
MCMC model composition \citep{MadiganYork1995} approach: Starting from the null
model, we move through $\Gamma$ by successive slight modifications of the
configuration $\gamma$. The modifications are accepted with MH acceptance
probabilities, which ensures that models with higher posterior probability are
more likely to be visited; see \citet{SabanesHeld2010} for details. For all four
approaches (F1, F2, F3 and EB), we ran this model sampler for one million
iterations. To get an idea of the computational complexity, note that on average
10.8 (F2) and
22.1 (EB)
models could be evaluated per second (on 2.8\,GHz CPUs). All computations have
been implemented in an \texttt{R}-package including an efficient \texttt{C++}
core for the MCMC parts, which is available from the first author.

For all four approaches Table~\ref{tab:Pima-fps-inclusion-probs} shows clear
evidence for the covariates $x_{2}, x_{5}, x_{6}$ and $x_{7}$ with posterior
inclusion probabilities over 99\%, while the other three covariates have
probabilities below 15\%. In comparison with the variable inclusion results for
the untransformed covariates in Table~\ref{tab:Pima-varselect-inclusion-probs},
it is interesting that $x_{1}$ is no longer important when FP transformations
are considered, while $x_{7}$ is much more important.

\begin{table}
  \centering
%
\begin{tabular}{lrrrr}\toprule
\multicolumn{1}{l}{}&\multicolumn{1}{c}{F1}&\multicolumn{1}{c}{F2}&\multicolumn{1}{c}{F3}&\multicolumn{1}{c}{EB}\tabularnewline
\midrule
$x_{ 1 }$&$0.119$&$0.125$&$0.135$&$0.144$\tabularnewline
$x_{ 2 }$&$1.000$&$1.000$&$1.000$&$1.000$\tabularnewline
$x_{ 3 }$&$0.050$&$0.052$&$0.054$&$0.054$\tabularnewline
$x_{ 4 }$&$0.032$&$0.033$&$0.033$&$0.035$\tabularnewline
$x_{ 5 }$&$0.999$&$0.999$&$0.999$&$0.999$\tabularnewline
$x_{ 6 }$&$0.992$&$0.993$&$0.993$&$0.994$\tabularnewline
$x_{ 7 }$&$0.999$&$0.999$&$0.999$&$0.999$\tabularnewline
\bottomrule
\end{tabular}\caption{Posterior probabilities for variable inclusion in the Pima data when FP
  transformations are considered. The probabilities are based on
  671\,525 (F1),
  719\,929 (F2),
  758\,616 (F3), and
  777\,531 (EB) visited models.}
\label{tab:Pima-fps-inclusion-probs}
\end{table}

\begin{figure}[htbp]
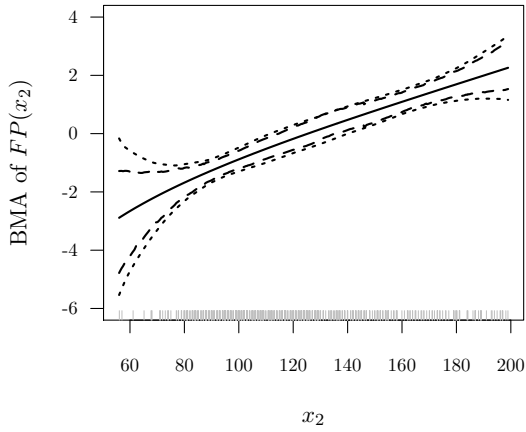
%
  \centering%
  \caption{Model-averaged FP transformations of selected Pima Indians covariates
    under hyperprior F1. Means (solid lines), pointwise (dashed lines) as
    well as simultaneous (dotted lines) 95\% credible intervals are given.
    Small ticks above the x-axes indicate data locations.}%
  \label{fig:pima-fps-bma-f1}%
\subfloat[Covariate $x_{2}$ (plasma glucose concentration)%
\label{fig:pima-fps-bma-f1:x2}]{%
\input{out-pima-fps-bma-f1-x2.tikz}
}%
\subfloat[Covariate $x_{5}$ (BMI)%
\label{fig:pima-fps-bma-f1:x5}]{%
\input{out-pima-fps-bma-f1-x5.tikz}
}%
\\ 
\subfloat[Covariate $x_{6}$ (diabetes pedigree function)%
\label{fig:pima-fps-bma-f1:x6}]{%
\input{out-pima-fps-bma-f1-x6.tikz}
}%
\subfloat[Covariate $x_{7}$ (age)%
\label{fig:pima-fps-bma-f1:x7}]{%
\input{out-pima-fps-bma-f1-x7.tikz}
}%
\end{figure}

In addition to examining the marginal inclusion probabilities, it is necessary
to look at the transformations of the covariates. Since all four approaches
produce similar variable inclusion probabilities and also share the MAP model
$\boldsymbol{x}_{i} = (x_{2i}, x_{5i}^{-2}, x_{6i}^{-1/2}, x_{7i}^{-2})^{T}$, we
only look in detail at the approach F1 (the three other producing again very
similar results). In order to account for the model uncertainty, it is best to
look at model-averaged estimates of variable transformations, conditional on
variable inclusion. To this end we varied the transformation of one of the
covariates $x_{2}, x_{5}, x_{6}, x_{7}$ while fixing the others at their MAP
configuration. Averaging over the 44 models each then results in the panels in
Figure~\ref{fig:pima-fps-bma-f1}.
Plasma glucose concentration ($x_{2}$) seems to have a strong positive linear
association with diabetes log-odds, while the estimated positive effect of BMI
($x_{5}$) is levelling off non-linearly for (rare) high values and is weaker
overall. Even smaller is the estimated positive effect of diabetes pedigree
function ($x_{6}$) with the largest increase in diabetes risk between
$x_{6}=0.1$ and $x_{6}=0.5$. The remaining estimated association of age
($x_{7}$) is clearly non-linear, with higher diabetes risk for middle-aged
participants.
These results are qualitatively similar to those obtained by
\citet[p.~665]{CottetKohnNott2008} for a larger subset of the original Pima
Indians data.

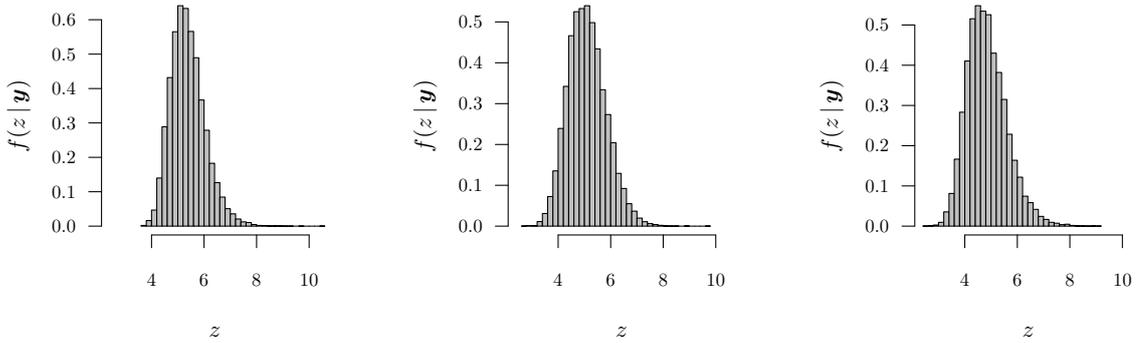
\begin{figure}
  \centering%
  \caption{Comparison of marginal posteriors for $z=\log(g)$ under priors F1, F2
    and F3. The histograms are based on the model average over the respective
    1000 models with highest posterior probability visited by the model
    samplers.}%
  \label{fig:pima-fps-marginal-z-posteriors}%
  \subfloat[Posterior for prior F1.%
  \label{fig:pima-fps-marginal-z-posteriors:f1}]{%
\begin{tikzpicture}[x=1pt,y=1pt]
\draw[color=white,opacity=0] (0,0) rectangle (144.54,144.54);
\begin{scope}
\path[clip] (  0.00,  0.00) rectangle (144.54,144.54);
\definecolor[named]{drawColor}{rgb}{0.03,0.65,0.25}
\definecolor[named]{fillColor}{rgb}{0.00,0.01,0.54}
\definecolor[named]{drawColor}{rgb}{0.00,0.00,0.00}

\node[color=drawColor,anchor=base,inner sep=0pt, outer sep=0pt, scale=  0.80] at ( 96.27, 10.56) {$z$};

\node[rotate= 90.00,color=drawColor,anchor=base,inner sep=0pt, outer sep=0pt, scale=  0.80] at ( 24.96, 93.87) {$f(z \given \boldsymbol{y})$};
\end{scope}
\begin{scope}
\path[clip] (  0.00,  0.00) rectangle (144.54,144.54);
\definecolor[named]{drawColor}{rgb}{0.03,0.65,0.25}
\definecolor[named]{fillColor}{rgb}{0.00,0.01,0.54}
\definecolor[named]{drawColor}{rgb}{0.00,0.00,0.00}
\definecolor[named]{fillColor}{rgb}{1.00,1.00,1.00}

\draw[color=drawColor,line cap=round,line join=round,fill=fillColor,] ( 72.48, 48.96) -- (131.48, 48.96);

\draw[color=drawColor,line cap=round,line join=round,fill=fillColor,] ( 72.48, 48.96) -- ( 72.48, 44.16);

\draw[color=drawColor,line cap=round,line join=round,fill=fillColor,] ( 92.15, 48.96) -- ( 92.15, 44.16);

\draw[color=drawColor,line cap=round,line join=round,fill=fillColor,] (111.81, 48.96) -- (111.81, 44.16);

\draw[color=drawColor,line cap=round,line join=round,fill=fillColor,] (131.48, 48.96) -- (131.48, 44.16);

\node[color=drawColor,anchor=base,inner sep=0pt, outer sep=0pt, scale=  0.64] at ( 72.48, 29.76) {4};

\node[color=drawColor,anchor=base,inner sep=0pt, outer sep=0pt, scale=  0.64] at ( 92.15, 29.76) {6};

\node[color=drawColor,anchor=base,inner sep=0pt, outer sep=0pt, scale=  0.64] at (111.81, 29.76) {8};

\node[color=drawColor,anchor=base,inner sep=0pt, outer sep=0pt, scale=  0.64] at (131.48, 29.76) {10};

\draw[color=drawColor,line cap=round,line join=round,fill=fillColor,] ( 53.76, 52.29) -- ( 53.76,130.16);

\draw[color=drawColor,line cap=round,line join=round,fill=fillColor,] ( 53.76, 52.29) -- ( 48.96, 52.29);

\draw[color=drawColor,line cap=round,line join=round,fill=fillColor,] ( 53.76, 65.27) -- ( 48.96, 65.27);

\draw[color=drawColor,line cap=round,line join=round,fill=fillColor,] ( 53.76, 78.25) -- ( 48.96, 78.25);

\draw[color=drawColor,line cap=round,line join=round,fill=fillColor,] ( 53.76, 91.23) -- ( 48.96, 91.23);

\draw[color=drawColor,line cap=round,line join=round,fill=fillColor,] ( 53.76,104.20) -- ( 48.96,104.20);

\draw[color=drawColor,line cap=round,line join=round,fill=fillColor,] ( 53.76,117.18) -- ( 48.96,117.18);

\draw[color=drawColor,line cap=round,line join=round,fill=fillColor,] ( 53.76,130.16) -- ( 48.96,130.16);

\node[color=drawColor,anchor=base east,inner sep=0pt, outer sep=0pt, scale=  0.64] at ( 44.16, 50.08) {0.0};

\node[color=drawColor,anchor=base east,inner sep=0pt, outer sep=0pt, scale=  0.64] at ( 44.16, 63.06) {0.1};

\node[color=drawColor,anchor=base east,inner sep=0pt, outer sep=0pt, scale=  0.64] at ( 44.16, 76.04) {0.2};

\node[color=drawColor,anchor=base east,inner sep=0pt, outer sep=0pt, scale=  0.64] at ( 44.16, 89.02) {0.3};

\node[color=drawColor,anchor=base east,inner sep=0pt, outer sep=0pt, scale=  0.64] at ( 44.16,102.00) {0.4};

\node[color=drawColor,anchor=base east,inner sep=0pt, outer sep=0pt, scale=  0.64] at ( 44.16,114.98) {0.5};

\node[color=drawColor,anchor=base east,inner sep=0pt, outer sep=0pt, scale=  0.64] at ( 44.16,127.96) {0.6};
\end{scope}
\begin{scope}
\path[clip] ( 53.76, 48.96) rectangle (138.78,138.78);
\definecolor[named]{drawColor}{rgb}{0.03,0.65,0.25}
\definecolor[named]{fillColor}{rgb}{0.00,0.01,0.54}
\definecolor[named]{drawColor}{rgb}{0.00,0.00,0.00}
\definecolor[named]{fillColor}{rgb}{0.75,0.75,0.75}

\draw[color=drawColor,line cap=round,line join=round,fill=fillColor,] ( 68.55, 52.29) rectangle ( 70.52, 52.55);

\draw[color=drawColor,line cap=round,line join=round,fill=fillColor,] ( 70.52, 52.29) rectangle ( 72.48, 54.40);

\draw[color=drawColor,line cap=round,line join=round,fill=fillColor,] ( 72.48, 52.29) rectangle ( 74.45, 58.35);

\draw[color=drawColor,line cap=round,line join=round,fill=fillColor,] ( 74.45, 52.29) rectangle ( 76.42, 70.43);

\draw[color=drawColor,line cap=round,line join=round,fill=fillColor,] ( 76.42, 52.29) rectangle ( 78.38, 89.77);

\draw[color=drawColor,line cap=round,line join=round,fill=fillColor,] ( 78.38, 52.29) rectangle ( 80.35,108.33);

\draw[color=drawColor,line cap=round,line join=round,fill=fillColor,] ( 80.35, 52.29) rectangle ( 82.32,125.62);

\draw[color=drawColor,line cap=round,line join=round,fill=fillColor,] ( 82.32, 52.29) rectangle ( 84.28,135.45);

\draw[color=drawColor,line cap=round,line join=round,fill=fillColor,] ( 84.28, 52.29) rectangle ( 86.25,134.41);

\draw[color=drawColor,line cap=round,line join=round,fill=fillColor,] ( 86.25, 52.29) rectangle ( 88.22,125.75);

\draw[color=drawColor,line cap=round,line join=round,fill=fillColor,] ( 88.22, 52.29) rectangle ( 90.18,115.82);

\draw[color=drawColor,line cap=round,line join=round,fill=fillColor,] ( 90.18, 52.29) rectangle ( 92.15, 99.89);

\draw[color=drawColor,line cap=round,line join=round,fill=fillColor,] ( 92.15, 52.29) rectangle ( 94.12, 88.47);

\draw[color=drawColor,line cap=round,line join=round,fill=fillColor,] ( 94.12, 52.29) rectangle ( 96.08, 76.01);

\draw[color=drawColor,line cap=round,line join=round,fill=fillColor,] ( 96.08, 52.29) rectangle ( 98.05, 68.71);

\draw[color=drawColor,line cap=round,line join=round,fill=fillColor,] ( 98.05, 52.29) rectangle (100.02, 63.25);

\draw[color=drawColor,line cap=round,line join=round,fill=fillColor,] (100.02, 52.29) rectangle (101.98, 58.87);

\draw[color=drawColor,line cap=round,line join=round,fill=fillColor,] (101.98, 52.29) rectangle (103.95, 56.93);

\draw[color=drawColor,line cap=round,line join=round,fill=fillColor,] (103.95, 52.29) rectangle (105.92, 54.98);

\draw[color=drawColor,line cap=round,line join=round,fill=fillColor,] (105.92, 52.29) rectangle (107.88, 53.91);

\draw[color=drawColor,line cap=round,line join=round,fill=fillColor,] (107.88, 52.29) rectangle (109.85, 53.65);

\draw[color=drawColor,line cap=round,line join=round,fill=fillColor,] (109.85, 52.29) rectangle (111.81, 52.90);

\draw[color=drawColor,line cap=round,line join=round,fill=fillColor,] (111.81, 52.29) rectangle (113.78, 52.55);

\draw[color=drawColor,line cap=round,line join=round,fill=fillColor,] (113.78, 52.29) rectangle (115.75, 52.51);

\draw[color=drawColor,line cap=round,line join=round,fill=fillColor,] (115.75, 52.29) rectangle (117.71, 52.35);

\draw[color=drawColor,line cap=round,line join=round,fill=fillColor,] (117.71, 52.29) rectangle (119.68, 52.38);

\draw[color=drawColor,line cap=round,line join=round,fill=fillColor,] (119.68, 52.29) rectangle (121.65, 52.35);

\draw[color=drawColor,line cap=round,line join=round,fill=fillColor,] (121.65, 52.29) rectangle (123.61, 52.32);

\draw[color=drawColor,line cap=round,line join=round,fill=fillColor,] (123.61, 52.29) rectangle (125.58, 52.32);

\draw[color=drawColor,line cap=round,line join=round,fill=fillColor,] (125.58, 52.29) rectangle (127.55, 52.29);

\draw[color=drawColor,line cap=round,line join=round,fill=fillColor,] (127.55, 52.29) rectangle (129.51, 52.32);

\draw[color=drawColor,line cap=round,line join=round,fill=fillColor,] (129.51, 52.29) rectangle (131.48, 52.29);

\draw[color=drawColor,line cap=round,line join=round,fill=fillColor,] (131.48, 52.29) rectangle (133.45, 52.29);

\draw[color=drawColor,line cap=round,line join=round,fill=fillColor,] (133.45, 52.29) rectangle (135.41, 52.29);

\draw[color=drawColor,line cap=round,line join=round,fill=fillColor,] (135.41, 52.29) rectangle (137.38, 52.32);
\end{scope}
\end{tikzpicture}
  }%
  \subfloat[Posterior for prior F2.%
  \label{fig:pima-fps-marginal-z-posteriors:f2}]{%
\begin{tikzpicture}[x=1pt,y=1pt]
\draw[color=white,opacity=0] (0,0) rectangle (144.54,144.54);
\begin{scope}
\path[clip] (  0.00,  0.00) rectangle (144.54,144.54);
\definecolor[named]{drawColor}{rgb}{0.94,0.82,0.13}
\definecolor[named]{fillColor}{rgb}{0.47,0.25,0.19}
\definecolor[named]{drawColor}{rgb}{0.00,0.00,0.00}

\node[color=drawColor,anchor=base,inner sep=0pt, outer sep=0pt, scale=  0.80] at ( 96.27, 10.56) {$z$};

\node[rotate= 90.00,color=drawColor,anchor=base,inner sep=0pt, outer sep=0pt, scale=  0.80] at ( 24.96, 93.87) {$f(z \given \boldsymbol{y})$};
\end{scope}
\begin{scope}
\path[clip] (  0.00,  0.00) rectangle (144.54,144.54);
\definecolor[named]{drawColor}{rgb}{0.94,0.82,0.13}
\definecolor[named]{fillColor}{rgb}{0.47,0.25,0.19}
\definecolor[named]{drawColor}{rgb}{0.00,0.00,0.00}
\definecolor[named]{fillColor}{rgb}{1.00,1.00,1.00}

\draw[color=drawColor,line cap=round,line join=round,fill=fillColor,] ( 72.48, 48.96) -- (131.48, 48.96);

\draw[color=drawColor,line cap=round,line join=round,fill=fillColor,] ( 72.48, 48.96) -- ( 72.48, 44.16);

\draw[color=drawColor,line cap=round,line join=round,fill=fillColor,] ( 92.15, 48.96) -- ( 92.15, 44.16);

\draw[color=drawColor,line cap=round,line join=round,fill=fillColor,] (111.81, 48.96) -- (111.81, 44.16);

\draw[color=drawColor,line cap=round,line join=round,fill=fillColor,] (131.48, 48.96) -- (131.48, 44.16);

\node[color=drawColor,anchor=base,inner sep=0pt, outer sep=0pt, scale=  0.64] at ( 72.48, 29.76) {4};

\node[color=drawColor,anchor=base,inner sep=0pt, outer sep=0pt, scale=  0.64] at ( 92.15, 29.76) {6};

\node[color=drawColor,anchor=base,inner sep=0pt, outer sep=0pt, scale=  0.64] at (111.81, 29.76) {8};

\node[color=drawColor,anchor=base,inner sep=0pt, outer sep=0pt, scale=  0.64] at (131.48, 29.76) {10};

\draw[color=drawColor,line cap=round,line join=round,fill=fillColor,] ( 53.76, 52.29) -- ( 53.76,129.33);

\draw[color=drawColor,line cap=round,line join=round,fill=fillColor,] ( 53.76, 52.29) -- ( 48.96, 52.29);

\draw[color=drawColor,line cap=round,line join=round,fill=fillColor,] ( 53.76, 67.70) -- ( 48.96, 67.70);

\draw[color=drawColor,line cap=round,line join=round,fill=fillColor,] ( 53.76, 83.10) -- ( 48.96, 83.10);

\draw[color=drawColor,line cap=round,line join=round,fill=fillColor,] ( 53.76, 98.51) -- ( 48.96, 98.51);

\draw[color=drawColor,line cap=round,line join=round,fill=fillColor,] ( 53.76,113.92) -- ( 48.96,113.92);

\draw[color=drawColor,line cap=round,line join=round,fill=fillColor,] ( 53.76,129.33) -- ( 48.96,129.33);

\node[color=drawColor,anchor=base east,inner sep=0pt, outer sep=0pt, scale=  0.64] at ( 44.16, 50.08) {0.0};

\node[color=drawColor,anchor=base east,inner sep=0pt, outer sep=0pt, scale=  0.64] at ( 44.16, 65.49) {0.1};

\node[color=drawColor,anchor=base east,inner sep=0pt, outer sep=0pt, scale=  0.64] at ( 44.16, 80.90) {0.2};

\node[color=drawColor,anchor=base east,inner sep=0pt, outer sep=0pt, scale=  0.64] at ( 44.16, 96.31) {0.3};

\node[color=drawColor,anchor=base east,inner sep=0pt, outer sep=0pt, scale=  0.64] at ( 44.16,111.72) {0.4};

\node[color=drawColor,anchor=base east,inner sep=0pt, outer sep=0pt, scale=  0.64] at ( 44.16,127.12) {0.5};
\end{scope}
\begin{scope}
\path[clip] ( 53.76, 48.96) rectangle (138.78,138.78);
\definecolor[named]{drawColor}{rgb}{0.94,0.82,0.13}
\definecolor[named]{fillColor}{rgb}{0.47,0.25,0.19}
\definecolor[named]{drawColor}{rgb}{0.00,0.00,0.00}
\definecolor[named]{fillColor}{rgb}{0.75,0.75,0.75}

\draw[color=drawColor,line cap=round,line join=round,fill=fillColor,] ( 58.72, 52.29) rectangle ( 60.68, 52.33);

\draw[color=drawColor,line cap=round,line join=round,fill=fillColor,] ( 60.68, 52.29) rectangle ( 62.65, 52.52);

\draw[color=drawColor,line cap=round,line join=round,fill=fillColor,] ( 62.65, 52.29) rectangle ( 64.62, 52.48);

\draw[color=drawColor,line cap=round,line join=round,fill=fillColor,] ( 64.62, 52.29) rectangle ( 66.58, 54.02);

\draw[color=drawColor,line cap=round,line join=round,fill=fillColor,] ( 66.58, 52.29) rectangle ( 68.55, 57.06);

\draw[color=drawColor,line cap=round,line join=round,fill=fillColor,] ( 68.55, 52.29) rectangle ( 70.52, 63.46);

\draw[color=drawColor,line cap=round,line join=round,fill=fillColor,] ( 70.52, 52.29) rectangle ( 72.48, 73.13);

\draw[color=drawColor,line cap=round,line join=round,fill=fillColor,] ( 72.48, 52.29) rectangle ( 74.45, 89.15);

\draw[color=drawColor,line cap=round,line join=round,fill=fillColor,] ( 74.45, 52.29) rectangle ( 76.42,105.02);

\draw[color=drawColor,line cap=round,line join=round,fill=fillColor,] ( 76.42, 52.29) rectangle ( 78.38,124.13);

\draw[color=drawColor,line cap=round,line join=round,fill=fillColor,] ( 78.38, 52.29) rectangle ( 80.35,133.18);

\draw[color=drawColor,line cap=round,line join=round,fill=fillColor,] ( 80.35, 52.29) rectangle ( 82.32,134.37);

\draw[color=drawColor,line cap=round,line join=round,fill=fillColor,] ( 82.32, 52.29) rectangle ( 84.28,135.45);

\draw[color=drawColor,line cap=round,line join=round,fill=fillColor,] ( 84.28, 52.29) rectangle ( 86.25,129.06);

\draw[color=drawColor,line cap=round,line join=round,fill=fillColor,] ( 86.25, 52.29) rectangle ( 88.22,119.16);

\draw[color=drawColor,line cap=round,line join=round,fill=fillColor,] ( 88.22, 52.29) rectangle ( 90.18,103.75);

\draw[color=drawColor,line cap=round,line join=round,fill=fillColor,] ( 90.18, 52.29) rectangle ( 92.15, 94.39);

\draw[color=drawColor,line cap=round,line join=round,fill=fillColor,] ( 92.15, 52.29) rectangle ( 94.12, 83.76);

\draw[color=drawColor,line cap=round,line join=round,fill=fillColor,] ( 94.12, 52.29) rectangle ( 96.08, 72.20);

\draw[color=drawColor,line cap=round,line join=round,fill=fillColor,] ( 96.08, 52.29) rectangle ( 98.05, 66.50);

\draw[color=drawColor,line cap=round,line join=round,fill=fillColor,] ( 98.05, 52.29) rectangle (100.02, 60.76);

\draw[color=drawColor,line cap=round,line join=round,fill=fillColor,] (100.02, 52.29) rectangle (101.98, 57.95);

\draw[color=drawColor,line cap=round,line join=round,fill=fillColor,] (101.98, 52.29) rectangle (103.95, 55.33);

\draw[color=drawColor,line cap=round,line join=round,fill=fillColor,] (103.95, 52.29) rectangle (105.92, 53.98);

\draw[color=drawColor,line cap=round,line join=round,fill=fillColor,] (105.92, 52.29) rectangle (107.88, 53.25);

\draw[color=drawColor,line cap=round,line join=round,fill=fillColor,] (107.88, 52.29) rectangle (109.85, 52.90);

\draw[color=drawColor,line cap=round,line join=round,fill=fillColor,] (109.85, 52.29) rectangle (111.81, 52.59);

\draw[color=drawColor,line cap=round,line join=round,fill=fillColor,] (111.81, 52.29) rectangle (113.78, 52.36);

\draw[color=drawColor,line cap=round,line join=round,fill=fillColor,] (113.78, 52.29) rectangle (115.75, 52.33);

\draw[color=drawColor,line cap=round,line join=round,fill=fillColor,] (115.75, 52.29) rectangle (117.71, 52.33);

\draw[color=drawColor,line cap=round,line join=round,fill=fillColor,] (117.71, 52.29) rectangle (119.68, 52.29);

\draw[color=drawColor,line cap=round,line join=round,fill=fillColor,] (119.68, 52.29) rectangle (121.65, 52.36);

\draw[color=drawColor,line cap=round,line join=round,fill=fillColor,] (121.65, 52.29) rectangle (123.61, 52.29);

\draw[color=drawColor,line cap=round,line join=round,fill=fillColor,] (123.61, 52.29) rectangle (125.58, 52.29);

\draw[color=drawColor,line cap=round,line join=round,fill=fillColor,] (125.58, 52.29) rectangle (127.55, 52.29);

\draw[color=drawColor,line cap=round,line join=round,fill=fillColor,] (127.55, 52.29) rectangle (129.51, 52.33);
\end{scope}
\end{tikzpicture}
  }%
  \subfloat[Posterior for prior F3.%
  \label{fig:pima-fps-marginal-z-posteriors:f3}]{%
\begin{tikzpicture}[x=1pt,y=1pt]
\draw[color=white,opacity=0] (0,0) rectangle (144.54,144.54);
\begin{scope}
\path[clip] (  0.00,  0.00) rectangle (144.54,144.54);
\definecolor[named]{drawColor}{rgb}{0.09,0.12,0.28}
\definecolor[named]{fillColor}{rgb}{0.06,0.62,1.00}
\definecolor[named]{drawColor}{rgb}{0.00,0.00,0.00}

\node[color=drawColor,anchor=base,inner sep=0pt, outer sep=0pt, scale=  0.80] at ( 96.27, 10.56) {$z$};

\node[rotate= 90.00,color=drawColor,anchor=base,inner sep=0pt, outer sep=0pt, scale=  0.80] at ( 24.96, 93.87) {$f(z \given \boldsymbol{y})$};
\end{scope}
\begin{scope}
\path[clip] (  0.00,  0.00) rectangle (144.54,144.54);
\definecolor[named]{drawColor}{rgb}{0.09,0.12,0.28}
\definecolor[named]{fillColor}{rgb}{0.06,0.62,1.00}
\definecolor[named]{drawColor}{rgb}{0.00,0.00,0.00}
\definecolor[named]{fillColor}{rgb}{1.00,1.00,1.00}

\draw[color=drawColor,line cap=round,line join=round,fill=fillColor,] ( 72.48, 48.96) -- (131.48, 48.96);

\draw[color=drawColor,line cap=round,line join=round,fill=fillColor,] ( 72.48, 48.96) -- ( 72.48, 44.16);

\draw[color=drawColor,line cap=round,line join=round,fill=fillColor,] ( 92.15, 48.96) -- ( 92.15, 44.16);

\draw[color=drawColor,line cap=round,line join=round,fill=fillColor,] (111.81, 48.96) -- (111.81, 44.16);

\draw[color=drawColor,line cap=round,line join=round,fill=fillColor,] (131.48, 48.96) -- (131.48, 44.16);

\node[color=drawColor,anchor=base,inner sep=0pt, outer sep=0pt, scale=  0.64] at ( 72.48, 29.76) {4};

\node[color=drawColor,anchor=base,inner sep=0pt, outer sep=0pt, scale=  0.64] at ( 92.15, 29.76) {6};

\node[color=drawColor,anchor=base,inner sep=0pt, outer sep=0pt, scale=  0.64] at (111.81, 29.76) {8};

\node[color=drawColor,anchor=base,inner sep=0pt, outer sep=0pt, scale=  0.64] at (131.48, 29.76) {10};

\draw[color=drawColor,line cap=round,line join=round,fill=fillColor,] ( 53.76, 52.29) -- ( 53.76,128.17);

\draw[color=drawColor,line cap=round,line join=round,fill=fillColor,] ( 53.76, 52.29) -- ( 48.96, 52.29);

\draw[color=drawColor,line cap=round,line join=round,fill=fillColor,] ( 53.76, 67.46) -- ( 48.96, 67.46);

\draw[color=drawColor,line cap=round,line join=round,fill=fillColor,] ( 53.76, 82.64) -- ( 48.96, 82.64);

\draw[color=drawColor,line cap=round,line join=round,fill=fillColor,] ( 53.76, 97.82) -- ( 48.96, 97.82);

\draw[color=drawColor,line cap=round,line join=round,fill=fillColor,] ( 53.76,112.99) -- ( 48.96,112.99);

\draw[color=drawColor,line cap=round,line join=round,fill=fillColor,] ( 53.76,128.17) -- ( 48.96,128.17);

\node[color=drawColor,anchor=base east,inner sep=0pt, outer sep=0pt, scale=  0.64] at ( 44.16, 50.08) {0.0};

\node[color=drawColor,anchor=base east,inner sep=0pt, outer sep=0pt, scale=  0.64] at ( 44.16, 65.26) {0.1};

\node[color=drawColor,anchor=base east,inner sep=0pt, outer sep=0pt, scale=  0.64] at ( 44.16, 80.44) {0.2};

\node[color=drawColor,anchor=base east,inner sep=0pt, outer sep=0pt, scale=  0.64] at ( 44.16, 95.61) {0.3};

\node[color=drawColor,anchor=base east,inner sep=0pt, outer sep=0pt, scale=  0.64] at ( 44.16,110.79) {0.4};

\node[color=drawColor,anchor=base east,inner sep=0pt, outer sep=0pt, scale=  0.64] at ( 44.16,125.96) {0.5};
\end{scope}
\begin{scope}
\path[clip] ( 53.76, 48.96) rectangle (138.78,138.78);
\definecolor[named]{drawColor}{rgb}{0.09,0.12,0.28}
\definecolor[named]{fillColor}{rgb}{0.06,0.62,1.00}
\definecolor[named]{drawColor}{rgb}{0.00,0.00,0.00}
\definecolor[named]{fillColor}{rgb}{0.75,0.75,0.75}

\draw[color=drawColor,line cap=round,line join=round,fill=fillColor,] ( 56.75, 52.29) rectangle ( 58.72, 52.32);

\draw[color=drawColor,line cap=round,line join=round,fill=fillColor,] ( 58.72, 52.29) rectangle ( 60.68, 52.44);

\draw[color=drawColor,line cap=round,line join=round,fill=fillColor,] ( 60.68, 52.29) rectangle ( 62.65, 52.70);

\draw[color=drawColor,line cap=round,line join=round,fill=fillColor,] ( 62.65, 52.29) rectangle ( 64.62, 53.73);

\draw[color=drawColor,line cap=round,line join=round,fill=fillColor,] ( 64.62, 52.29) rectangle ( 66.58, 57.67);

\draw[color=drawColor,line cap=round,line join=round,fill=fillColor,] ( 66.58, 52.29) rectangle ( 68.55, 64.62);

\draw[color=drawColor,line cap=round,line join=round,fill=fillColor,] ( 68.55, 52.29) rectangle ( 70.52, 77.56);

\draw[color=drawColor,line cap=round,line join=round,fill=fillColor,] ( 70.52, 52.29) rectangle ( 72.48, 95.27);

\draw[color=drawColor,line cap=round,line join=round,fill=fillColor,] ( 72.48, 52.29) rectangle ( 74.45,114.59);

\draw[color=drawColor,line cap=round,line join=round,fill=fillColor,] ( 74.45, 52.29) rectangle ( 76.42,130.52);

\draw[color=drawColor,line cap=round,line join=round,fill=fillColor,] ( 76.42, 52.29) rectangle ( 78.38,135.45);

\draw[color=drawColor,line cap=round,line join=round,fill=fillColor,] ( 78.38, 52.29) rectangle ( 80.35,133.40);

\draw[color=drawColor,line cap=round,line join=round,fill=fillColor,] ( 80.35, 52.29) rectangle ( 82.32,132.04);

\draw[color=drawColor,line cap=round,line join=round,fill=fillColor,] ( 82.32, 52.29) rectangle ( 84.28,117.55);

\draw[color=drawColor,line cap=round,line join=round,fill=fillColor,] ( 84.28, 52.29) rectangle ( 86.25,110.26);

\draw[color=drawColor,line cap=round,line join=round,fill=fillColor,] ( 86.25, 52.29) rectangle ( 88.22,100.09);

\draw[color=drawColor,line cap=round,line join=round,fill=fillColor,] ( 88.22, 52.29) rectangle ( 90.18, 86.96);

\draw[color=drawColor,line cap=round,line join=round,fill=fillColor,] ( 90.18, 52.29) rectangle ( 92.15, 77.14);

\draw[color=drawColor,line cap=round,line join=round,fill=fillColor,] ( 92.15, 52.29) rectangle ( 94.12, 70.76);

\draw[color=drawColor,line cap=round,line join=round,fill=fillColor,] ( 94.12, 52.29) rectangle ( 96.08, 63.59);

\draw[color=drawColor,line cap=round,line join=round,fill=fillColor,] ( 96.08, 52.29) rectangle ( 98.05, 61.16);

\draw[color=drawColor,line cap=round,line join=round,fill=fillColor,] ( 98.05, 52.29) rectangle (100.02, 58.62);

\draw[color=drawColor,line cap=round,line join=round,fill=fillColor,] (100.02, 52.29) rectangle (101.98, 55.97);

\draw[color=drawColor,line cap=round,line join=round,fill=fillColor,] (101.98, 52.29) rectangle (103.95, 54.83);

\draw[color=drawColor,line cap=round,line join=round,fill=fillColor,] (103.95, 52.29) rectangle (105.92, 53.73);

\draw[color=drawColor,line cap=round,line join=round,fill=fillColor,] (105.92, 52.29) rectangle (107.88, 53.39);

\draw[color=drawColor,line cap=round,line join=round,fill=fillColor,] (107.88, 52.29) rectangle (109.85, 52.86);

\draw[color=drawColor,line cap=round,line join=round,fill=fillColor,] (109.85, 52.29) rectangle (111.81, 53.01);

\draw[color=drawColor,line cap=round,line join=round,fill=fillColor,] (111.81, 52.29) rectangle (113.78, 52.55);

\draw[color=drawColor,line cap=round,line join=round,fill=fillColor,] (113.78, 52.29) rectangle (115.75, 52.40);

\draw[color=drawColor,line cap=round,line join=round,fill=fillColor,] (115.75, 52.29) rectangle (117.71, 52.36);

\draw[color=drawColor,line cap=round,line join=round,fill=fillColor,] (117.71, 52.29) rectangle (119.68, 52.32);

\draw[color=drawColor,line cap=round,line join=round,fill=fillColor,] (119.68, 52.29) rectangle (121.65, 52.36);

\draw[color=drawColor,line cap=round,line join=round,fill=fillColor,] (121.65, 52.29) rectangle (123.61, 52.32);
\end{scope}
\end{tikzpicture}
  }%
\end{figure}

The marginal posterior distributions for the covariance factor $g$ differ
slightly between the three prior choices F1, F2 and F3. Averaging over the best
1000 models in terms of posterior probability which have been visited by the
model sampler, we get the histograms (for $z=\log(g)$) in
Figure~\ref{fig:pima-fps-marginal-z-posteriors}. The corresponding posterior
means $\E(g \given \boldsymbol{y})$ decrease from 282.5
for F1, 219.2 for F2 to 179.1
for F3, and this trend is also visible in the histograms. The results suggest a
stronger prior shrinkage of the regression coefficients than that proposed by
the unit information prior's fixed value $g=n=532$ (\cf
Section~\ref{sec:generalized-hyper-g-prior:comparison}), as $\P(g < n \given
\boldsymbol{y})$ ranges from 90.9\% for F1 to
95.7\% for F3.

\section{Discussion}
\label{sec:discussion}

In this article, we presented a generalization of the $g$-prior to GLMs, which
can be interpreted analogously to the classical $g$-prior for normal linear
models. In our implementation, the shrinkage-controlling hyperparameter $g$ can
be assigned any hyperprior, thus giving rise to a large class of generalized
hyper-$g$ priors. For mixtures of classical $g$-priors,
\citet{LiangPauloMolinaClydeBerger2008} could investigate theoretical model
selection and prediction consistency properties. It would be desirable to also
investigate such properties for our generalized hyper-$g$ prior class. However,
as fewer closed form expressions are available, derivation of comparable proofs
will be more difficult in the GLM family.

Another important area of future research is the thorough comparison of the
generalized hyper-$g$ prior with the other approaches in the literature
summarized in Section~\ref{sec:generalized-hyper-g-prior:comparison}. For
example, exhaustive simulation studies could shed light on different
performances of the priors in variable selection. Perhaps also theoretical
results can be derived to explain the different properties of the approaches.

Bayesian inference for FPs in GLMs was in fact the motivating application for
this work. With huge model spaces to explore, the accurate numerical marginal
likelihood approximation is vital for this and similar typical applications of
the generalized hyper-$g$ prior. Alternative MCMC estimates of the marginal
likelihood were used to demonstrate the very good accuracy of the ILA. Yet, MCMC
would not be suited for replacing the deterministic ILA approach in the
stochastic model search, because the computation is slower by orders of
magnitude and would require careful automatic monitoring of convergence. Of
course, the deterministic marginal likelihood approximation could be used for
any type of stochastic model search, such as those recently proposed by
\citet{HansDobraWest2007} and \citet{Dobra2009}. 

Finally, we note that the classical $g$-prior has recently been extended in
other directions as well. In the context of supervised machine learning,
\citet{ZhangJordanYeung2008} replace
$\boldsymbol{X}_{\gamma}^{T}\boldsymbol{X}_{\gamma}$ by a (possibly singular)
kernel matrix $\boldsymbol{K}_{\gamma}$ and prove consistency properties for the
normal linear model. \citet{MaruyamaGeorge2008} remove the restriction of
$p_{\gamma} \leq n - 1$ for normal linear models by working with the singular
value decomposition (SVD) of the design matrix $\boldsymbol{X}_{\gamma}$. A
similar extension is the ``generalised singular $g$-prior'' defined by
\citet{West2003} in the factor regression context. Along these lines, our
generalized hyper-$g$ prior could also be extended to the $p_{\gamma} > n$ case
via the SVD $\boldsymbol{W}^{1/2}\boldsymbol{X}_{\gamma} =
\boldsymbol{U}_{\gamma}\boldsymbol{D}_{\gamma}\boldsymbol{V}_{\gamma}$. We could
just use the latent parameter $\boldsymbol{\delta}_{\gamma} =
\boldsymbol{V}_{\gamma}\boldsymbol{\beta}_{\gamma}$ of reduced dimension
$k_{\gamma} = n - 1$ instead of $\boldsymbol{\beta}_{\gamma} =
\boldsymbol{V}^{T}_{\gamma}\boldsymbol{\delta}_{\gamma}$. Defining the
corresponding design matrix as $\boldsymbol{Z}_{\gamma} =
\boldsymbol{W}^{-1/2}\boldsymbol{U}_{\gamma}\boldsymbol{D}_{\gamma}$, we have
$\boldsymbol{X}_{\gamma}\boldsymbol{\beta}_{\gamma} =
\boldsymbol{Z}_{\gamma}\boldsymbol{\delta}_{\gamma}$ and retain
$\boldsymbol{Z}_{\gamma}^{T}\boldsymbol{1}_{n} = \boldsymbol{0}_{k_{\gamma}}$.
Assigning the prior distribution $\boldsymbol{\delta}_{\gamma} \sim
\Nor_{k_{\gamma}}(\boldsymbol{0}_{k_{\gamma}}, g\phi
c\boldsymbol{D}_{\gamma}^{-2})$ then induces a normal prior on
$\boldsymbol{\beta}_{\gamma}$ with mean zero and singular precision
$(g\phi
c)^{-1}\boldsymbol{X}_{\gamma}^{T}\boldsymbol{W}\boldsymbol{X}_{\gamma}$, and
thus directly generalizes~\eqref{eq:generalized-g-prior}. Investigation of this
approach for GLMs with many covariates is another possibility for future
research.





\appendix{}

\section{Proof of prior mode zero}
\label{sec:prior-mode-zero-proof}

Consider the density function from~\eqref{eq:imag-posterior}. Dropping for
brevity the notational dependency on the model $\gamma$, it can be rewritten as
\begin{equation}
  \label{eq:imag-posterior-rewritten}
  f(\boldsymbol{\beta} \given g, \boldsymbol{y}_{0}) 
  \propto
  \exp
  \left\{
    \frac{1}{g\phi}
    \boldsymbol{w}^{T}
    \bigl(
    h(0) \boldsymbol{\theta} - b(\boldsymbol{\theta})
    \bigr)
  \right\},
\end{equation}
where $\boldsymbol{\theta} = (\theta_{1}, \dotsc, \theta_{n})^{T}$ and
$b(\boldsymbol{\theta}) = \bigl(b(\theta_{1}), \dotsc, b(\theta_{n})\bigr)^{T}$.
To prove that the mode is at $\boldsymbol{\beta} = \boldsymbol{0}_{p}$, note
that this is a solution of the score equation
\[
\dsepv{\boldsymbol{\beta}} 
\log f(\boldsymbol{\beta} \given g, \boldsymbol{y}_{0})  
=
\frac{1}{g\phi}
\left(
  h(0)
  \partialv{\boldsymbol{\theta}}{\boldsymbol{\beta}^{T}}
  -
  \partialv{b(\boldsymbol{\theta})}{\boldsymbol{\theta}^{T}}
  \partialv{\boldsymbol{\theta}}{\boldsymbol{\beta}^{T}}
\right)^{T}
\boldsymbol{w}
=
\boldsymbol{0}_{p},
\]
because $\boldsymbol{\beta} = \boldsymbol{0}_{p}$ implies that $b'(\theta_{i})
\equiv b'(\theta) = \mu = h(0)$ and hence
\[
\partialv{b(\boldsymbol{\theta})}{\boldsymbol{\theta}^{T}} =
\diag\bigl(b'(\theta_{1}), \dotsc, b'(\theta_{n})\bigr) =
h(0) \boldsymbol{I}_{n}.
\]

\section{Higher-order Laplace approximation}
\label{sec:higher-order-laplace}

Denote the standard Laplace approximation~\eqref{eq:cond-marg-lik-approximation}
by $\tilde{f}_{LA}(\boldsymbol{y} \given g, \gamma)$. Then
\citet[p.~148]{RaudenbushYangYosef2000} show that
\begin{equation}
  \label{eq:higher-order-laplace}
  f(\boldsymbol{y} \given g, \gamma)
  \approx
  \tilde{f}_{LA}(\boldsymbol{y} \given g, \gamma)
  \left[
    1 
    - \frac{1}{8} \sum_{i=1}^{n} d_{i}^{(3)} b_{i}^{2}
    - \frac{1}{48} \sum_{i=1}^{n} d_{i}^{(6)} b_{i}^{3}
    + \frac{5}{24} 
    \boldsymbol{k}^{T} (\boldsymbol{R}_{0\gamma}^{*})^{-1} \boldsymbol{k}  
  \right]
\end{equation}
is a sixth-order Laplace approximation when the canonical response function is
used. Here 
$d_{i}^{(m)} = d^{m}h/d\eta^{m}(\eta_{i}^{*})$
evaluated at
$\eta_{i}^{*} = \boldsymbol{x}_{0\gamma i}^{T}\boldsymbol{\beta}_{0\gamma}^{*}$,
$b_{i} = \boldsymbol{x}_{0\gamma i}^{T} (\boldsymbol{R}_{0\gamma}^{*})^{-1}
\boldsymbol{x}_{0\gamma i}$ and $\boldsymbol{k} = \sum_{i=1}^{n} d_{i}^{(2)}
b_{i} \boldsymbol{x}_{0\gamma i}$. Note that the quadratic forms can be
efficiently computed using the Cholesky decomposition
$\boldsymbol{R}_{0\gamma}^{*} = \boldsymbol{L}\boldsymbol{L}^{T}$, \eg
$\boldsymbol{k}^{T} (\boldsymbol{R}_{0\gamma}^{*})^{-1} \boldsymbol{k} =
\norm{\boldsymbol{v}}^{2}$ where $\boldsymbol{L}\boldsymbol{v} =
\boldsymbol{k}$.




\end{document}